\documentclass[superscriptaddress, showpacs, aps, twocolumn, pra, floatfix]{revtex4-1}

\usepackage{graphicx}
\usepackage{amsmath,amssymb,amsfonts}
\usepackage[percent]{overpic}
\usepackage{subfigure}
\usepackage{braket}
\usepackage{bm}
\usepackage[breaklinks=true,colorlinks,citecolor=blue,linkcolor=blue,urlcolor=blue]{hyperref}
\usepackage{comment}
\usepackage{cancel}
\usepackage{color}

\newcommand{\be}{\begin{equation}}
\newcommand{\ee}{\end{equation}}
\newcommand{\beq}{\begin{eqnarray}}
\newcommand{\eeq}{\end{eqnarray}}

\begin{document}
\title{Ultra-stable charging of fast-scrambling SYK quantum batteries}

\author{Dario Rosa}
\affiliation{School of Physics, Korea Institute for Advanced Study,
85 Hoegiro Dongdaemun-gu, Seoul 02455, Republic of Korea}
\email{Dario85@kias.re.kr}
\author{Davide Rossini}
\affiliation{Dipartimento di Fisica dell'Universit\`a di Pisa and INFN, Largo Bruno Pontecorvo 3, I-56127 Pisa, Italy}
\author{Gian Marcello Andolina}
\affiliation{NEST, Scuola Normale Superiore, I-56126 Pisa,~Italy}
\affiliation{Istituto Italiano di Tecnologia, Graphene Labs, Via Morego 30, I-16163 Genova,~Italy}
\author{Marco Polini}
\affiliation{{Dipartimento di Fisica dell'Universit\`a di Pisa, Largo Bruno Pontecorvo 3, I-56127 Pisa, Italy}}
\affiliation{{School of Physics \& Astronomy, University of Manchester, Oxford Road, Manchester M13 9PL, United Kingdom}}
\affiliation{{Istituto Italiano di Tecnologia, Graphene Labs, Via Morego 30, I-16163 Genova,~Italy}}
\author{Matteo Carrega}
\affiliation{NEST, Istituto Nanoscienze-CNR and Scuola Normale Superiore, I-56127 Pisa, Italy}

\begin{abstract}
Collective behavior strongly influences the charging dynamics of quantum batteries (QBs). Here, we study the impact of nonlocal correlations on the energy stored in a system of $N$ QBs.
A unitary charging protocol based on a Sachdev-Ye-Kitaev (SYK) quench Hamiltonian is thus introduced and analyzed. SYK models describe strongly interacting systems with nonlocal correlations and fast thermalization properties. Here, we demonstrate that, once charged, the average energy stored in the QB is very stable, realizing an ultraprecise charging protocol. By studying fluctuations of the average energy stored, we show that temporal fluctuations are strongly suppressed by the presence of nonlocal correlations at all time scales. A comparison with other paradigmatic examples of many-body QBs shows that this is linked to the collective dynamics of the SYK model and its high level of entanglement.
We argue that such feature relies on the fast scrambling property of the SYK Hamiltonian, and on its fast thermalization properties, promoting this as an ideal model for the ultimate temporal stability of a generic {QB}.
Finally, we show that the temporal evolution of the ergotropy, a quantity that  characterizes the amount of extractable work from a QB, can be a useful probe to infer the thermalization properties of a many-body quantum system.
\end{abstract}

\maketitle


\section{Introduction}
\label{sec:intro}

Recent advances in technological miniaturization and fabrication processes have led to the emergence of a new branch of research, dubbed ``quantum thermodynamics''~\cite{seifert2012, kosloff2013, esposito2009, campisi2011, levy2012, carrega2016}.
The study of thermodynamic concepts, such as work and heat, at the nanoscale, and the interplay with the laws of quantum mechanics, is crucial both from a fundamental and {a technological point} of view. 
A key goal here is to find new strategies to precisely control, store and manipulate work and energy~\cite{seifert2012, benenti2017}, with improved performances, eventually thanks to the presence of quantum coherence~\cite{goychuk2013, cavina2017, allahverdyan2013, brandner2018, agarwalla2018, coherence2018, carrega2019}.

\begin{figure}[ht]
  \includegraphics[width=8cm]{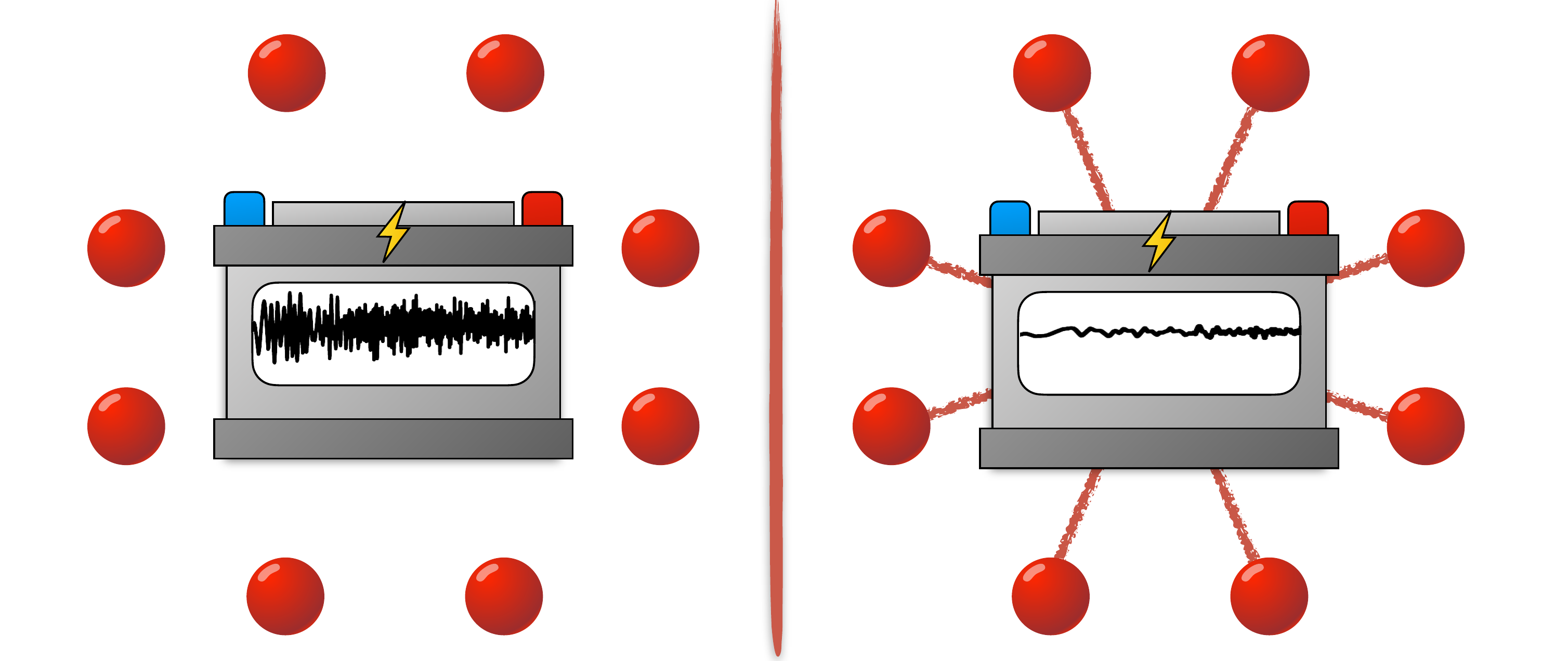}
  \caption{The charging protocol of a QB made of $N$ spin-1/2 cells.
    In the left panel, the battery is charged via a \emph{single-body} charging protocol. 
    The resulting charging protocol displays huge temporal fluctuations.
    By switching on a SYK-like quench  the charging protocol turns out to be \emph{collective} in nature, and intrinsic nonlocal correlations drastically suppress charging temporal fluctuations.}
  \label{fig0}
\end{figure}

In this framework, quantum batteries (QBs) have been introduced~\cite{alicki2013, campaioli2018} as small quantum systems able to temporarily store energy, to be used at a later stage. 
Different figures of merit, such as the charging time, the associated power, and the amount of extractable work, have been  analyzed~\cite{hovhannisyan2013, binder2015, campaioli2017, le2017, ferraro2017, andolina2018, zhang2018}  and bounds on their performances have been inspected, depending on the precise charging protocols~\cite{julia-farre2018, santos2019, garcia-pintos2019, andolina2019, friis2018}. 
These usually rely on an external charger, interacting with one or more cells of the QB~\cite{ferraro2017, andolina2018, barra2019}, or on unitary (local or global) evolution of the closed system in a nonequilibrium setting, {\it i.e.}~by exciting degrees of freedom with a quantum quench~\cite{campaioli2018, binder2015, campaioli2017, le2017, julia-farre2018, rossini2019}.

It has been shown that the presence of correlations and entanglement between quantum cells can have {a non-trivial} impact on both the charging power and the extractable work of QBs~\cite{binder2015,campaioli2017,julia-farre2018, andolina2019, farina2018}.
Indeed, part of the energy stored in the QB  can be locked by internal correlations, giving access only to a subset of the whole QB for useable work.
This has been quantified by introducing the so-called \emph{ergotropy}~\cite{allahverdyan2013, andolina2019}.
Although many-body effects can enhance charging performances~\cite{julia-farre2018, rossini2019}, strong and nonlocal correlations are required to achieve a true quantum advantage for QBs.
Recently, the role of random disorder and its impact on charging performances have been also inspected, showing that QBs exhibit typical behavior in the large $N$ limit given the spectral properties of the driving systems~\cite{gosh2019, caravelli2019}.

Moreover, after an initial growth, the average energy stored in a QB during the charging protocol inevitably undergoes fluctuations, which usually undermine its subsequent {use}.
It is thus of great importance to find protocols able to stabilize energy storage~\cite{santos2019, gherardini2019} or, even better, systems which intrinsically suppress these unwanted fluctuations.

In this work, we show that nonlocal and strongly chaotic correlations greatly help in improving charging stability of QBs, by suppressing temporal fluctuations associated to the average energy stored in a QB. 
To elucidate this point, we investigate a paradigmatic example of a strongly correlated {chaotic} system with nonlocal interactions, {\it i.e.}~we introduce and study QBs based on the so-called Sachdev-Ye-Kitaev (SYK) model.

SYK models~\cite{kitaev2015, sachdev1993, sachdev2010, maldacena2016, polchinski2016, franz2018} are currently receiving a lot of attention from very different communities. 
They describe strongly correlated quantum systems of (Majorana or Dirac) fermions with random all-to-all interactions. Interestingly, it has been shown~\cite{kitaev2015,maldacena2016,polchinski2016}, that, in the limit of large number of fermions,  many interesting quantities (such as the $n$-point correlation functions) can be exactly calculated.
Subsequently, the non-Fermi liquid behavior of SYK models has been studied~\cite{banerjee2016, lunkin2018, altland2019, kim:2019} and, in a completely different context, intriguing connections with black-hole physics and quantum gravity via holography have been explored~\cite{jensen2016, maldacena-stanford-yang2016, verlinde2016, maldacena2018}. 
Moreover, and in parallel, it has been shown that nonlocal correlations and random disorder result in highly chaotic dynamics, making these models extremely popular in the quantum chaos community too. 
The chaotic properties of the SYK models have been extensively investigated both from a random matrix theory point of view, starting from Ref.~\cite{garcia-garcia2016, cotler2016}, and by studying the so-called out-of-time-order correlators~\cite{kitaev2015, maldacena2016}.
The latter result in the saturation of the Maldacena-Shenker-Stanford bound on the Lyapunov exponent \cite{maldacena2015}, thus promoting the SYK models as concrete examples of systems satisfying the ``fast scrambling'' conjecture~\cite{sekino2008}, {\it i.e.} with the Liapunov exponent saturating  the bound $\lambda_L \leq 2\pi/\beta$, with $\beta$ denoting the inverse temperature.
In passing, we note that possible realizations of SYK have been proposed both in atomic~\cite{danshita2016, danshita2017} and solid-state physics~\cite{pikulin2018, alicea2017, chen2018}.

We consider {an} SYK system of $N$ QBs under {a} unitary charging protocol, as pictorially sketched in Fig.~\ref{fig0}. 
We argue that \emph{nonlocal} and \emph{chaotic} correlations, after the initial quench, lead to a very fast-- and homogeneous-- excitation of many energy levels, with huge creation of entanglement~\cite{liu2017, huang2019} and, more importantly, in a \emph{collective} {fashion}.
By characterizing fluctuations of the average energy stored, we show that this collective behavior is reflected in an exponential suppression of the temporal fluctuations at all time scales, leading to an ultra precise charging of the QB.
To corroborate these results we perform extensive numerical simulations, based on exact diagonalization, showing also a systematic comparison with a prototipical quantum system with many-body \emph{local} correlations, {\it i.e.}~a one-dimensional spin chain in the Anderson{, ergodic or} many-body localized (MBL) phase~\cite{kjall2014}.

This paper is organized as follows. 
In Sec.~\ref{sec:charging_generalities} we introduce the unitary charging protocol, its model-independent features and some preliminary definitions. 
In Sec.~\ref{sec:models} and Sec.~\ref{sec:interacting_models} we introduce the SYK model under investigation and we  compare its behavior  to  a spin chain in the MBL phase. 
We analyze energy fluctuations of different kinds, {\it i.e.} disorder, quantum, and temporal fluctuations, showing comparison between SYK and spin-chain (MBL or Anderson) based QBs.
In particular, we  demonstrate that SYK-based QBs result in exponentially suppressed \emph{temporal} fluctuations at all times, a peculiar feature that can be linked to the collective and nonlocal nature of the system and to its chaotic, and fast thermalizing, property.
{In Sec.~\ref{sec:quantum_chaos} we inspect the role played by quantum chaos in suppressing the temporal fluctuations. We will argue that the suppression of the temporal fluctuations of a generic, chaotic, QB can be linked to the spectral rigidity of the corresponding quench Hamiltonian.
On this respect, the very high degree of spectral rigidity of the SYK Hamiltonian, as observed in \cite{gharibyan2019}, explains the great performance of the corresponding SYK QBs.}
In Sec.~\ref{sec:ergotropy} {we make a digression and} we study the amount of extractable energy from a SYK-like QB, {\it i.e.}~its \emph{ergotropy}.
We  show that in general a SYK QB displays very low values of ergotropy, a feature that can be traced back to its highly entangling dynamics. Interestingly, we argue that by inspecting the time evolution of this quantity one can infer the thermalization time scale of a quantum system.
Sec.~\ref{sec:conclusions} contains a summary of our main findings.

\section{Charging protocol and energy fluctuations}
\label{sec:charging_generalities}

We study the charging mechanism of a QB, following a unitary protocol based on a double-sudden quench~\cite{campaioli2018, campaioli2017, binder2015, le2017}. 
The system is initially assumed to be in the ground state $\ket 0$ of a given time-independent Hamiltonian,
$\mathcal{\hat H}_0$ (empty battery). Subsequently, it evolves under the Hamiltonian
\be
\label{eq:charging_protocol_general}
\mathcal{\hat H}(t) = \mathcal{\hat H}_0 + \kappa \, \lambda(t) \, \mathcal{\hat H}_1 \ ,
\ee
where $\mathcal{\hat H}_1$ is a time-independent driving Hamiltonian and the dimensionless parameter $\kappa$
controls the relative strength between $\mathcal{\hat H}_0$ and $\mathcal{\hat H}_1$.
The function $\lambda(t)$ describes the charging time interval and is defined by 
\beq
\label{eq:lambda_def}
&& \lambda (t) = 0 \ , \qquad t < 0 \  \mathrm{and} \   t > \tau \ , \nonumber \\
&& \lambda (t) = 1 \ , \qquad 0< t < \tau \ , 
\eeq
with $\tau$ being the charging time.
Denoting {by} $\ket{\psi(t)}$ 
the evolved state under the total Hamiltonian~\eqref{eq:charging_protocol_general} (in this work we set $\hbar=1$),
the averaged energy stored in the QB at the end of the charging time is
\be
\label{eq:energy_stored}
E (\tau) \equiv \braket{\mathcal{\hat H}_0}_{\tau} - \braket{\mathcal{\hat H}_0}_{0} \ ,
\ee
where we defined $\braket{\mathcal{\hat H}_0}_{\tau} \equiv \bra{\psi(\tau)} \mathcal{\hat H}_0 \ket{\psi(\tau)}$.

Unless specified, we will always consider many-body QBs composed of $N$ cells (in our case $N$ qubits)
whose static Hamiltonian is given by
\begin{equation}
  \label{eq:QBham0}
  \mathcal{\hat H}_0 = h \, \sum_{j = 1}^N \hat \sigma_j^z ,
\end{equation}
$\hat \sigma_j^{\alpha}$ ($\alpha = x,y,z$) denoting the usual spin-$1/2$ Pauli operators corresponding to the $j$th qubit,
and $h$ being the QB energy scale.
We will also indicate with $\mathcal{\hat H}_0^{(M)} = h \, \sum_{j = 1}^M \hat \sigma_j^z$ a local portion of the QB,
once restricted to $M<N$ cells.

In a many-body QB, the stored energy $E(\tau)$ may display some universal features as a function of $\tau$:
it undergoes an initial growth for small $\tau$, while at larger times it fluctuates in time
around an average value~\cite{rossini2019}
\begin{equation}
  \bar E(\tau_1,\tau_2) = \frac{1}{\tau_2 - \tau_1}  \int_{\tau_1}^{\tau_2} \, d \tau \, \braket{\mathcal{\hat H}_0}_\tau  \ ,
\end{equation}
whose precise value depends on the specific model of QB considered.
In order to analyze the speed and performance of the charging protocol, we define an optimal charging time $\bar \tau$
as the one at which the energy stored in the QB reaches a value equal to a fixed fraction of the average energy.
Notice that, since temporal fluctuations are always present, the usual definition of $\bar \tau$ as the time at which
the energy stored in the battery reaches its maximum value is not well defined.

The charging precision of a QB is influenced by different and independent factors that may be responsible for temporal,
disorder and quantum fluctuations. The first kind of fluctuations can be quantified by computing
\begin{equation}
  \label{eq:temporal_variance}
  \big[ \sigma_N^{(\rm t)} (\tau_1,\tau_2) \big]^2 \equiv
  \big\langle\!\big\langle \bigg[ \int_{\tau_1}^{\tau_2} \, \frac{d \tau}{\tau_2 - \tau_1} \braket{\mathcal{\hat H}_0}_\tau^2  \bigg]
  - \bar E^2(\tau_1,\tau_2) \big\rangle\!\big\rangle \ .
\end{equation}
Hereafter we will use the symbol $\langle \! \langle \cdot \rangle \! \rangle$ in order to denote the average
over different realizations of the charging Hamiltonian $\mathcal{\hat H}_1$, which may depend on some parameters
drawn according to a given probability distribution.
We also define the dimensionless quantity 
\begin{equation}
  \label{eq:fluctuations_temp_def}
  \Sigma_N^{(\rm t)} (\tau_1,\tau_2) \equiv \frac{\sigma_N^{(\rm{t})} (\tau_1,\tau_2)}{\Delta_{\mathcal{\hat H}_0}/2} \ ,
\end{equation}
where $\Delta_{\mathcal{\hat H}_0} = Nh$ is the bandwidth of $\mathcal{\hat H}_0$ in Eq.~\eqref{eq:QBham0}.
More in general, the bandwidth of an Hermitian operator $\hat{\mathcal{O}}$ is defined as the norm
$\Delta_{\mathcal{\hat O}} \equiv \mu^{\rm(max)}_{\mathcal{\hat O}} - \mu^{\rm(min)}_{\mathcal{\hat O}}$,
where $\mu^{\rm(max)}_{\mathcal{\hat O}} (\mu^{\rm(min)}_{\mathcal{\hat O}})$ is its maximum (minimum) eigenvalue.

Disorder fluctuations may be responsible for an indetermination in $E(\tau)$ due to imperfections in the fabrication
of the QB, which can be modeled as suitable random parameters entering the full Hamiltonian $\mathcal{\hat H}$.
These fluctuations are defined by
\be
\label{eq:disorder_variance}
\big[ \sigma_N^{(\rm d)} (\tau)\big]^2 \equiv \big\langle\!\big\langle \braket{\mathcal{\hat H}_0}_\tau^2 \big\rangle\!\big\rangle
- \big\langle\!\big\langle \braket{\mathcal{\hat H}_0}_\tau \big\rangle\!\big\rangle^2 \ .
\ee
On the other hand, quantum fluctuations are caused by quantum indetermination, which is intrinsically present
in the charging process, since $\ket{\psi({\tau})}$ is not an eigenstate of $\mathcal{\hat H}_0$.
These can be quantified by 
\be
\label{eq:quantum_variance}
\big[ \sigma_N^{(\rm{q})} (\tau)\big]^2 \equiv \big\langle\!\big\langle \braket{\mathcal{\hat H}^2_0}_\tau -\braket{\mathcal{\hat H}_0}_\tau^2 \big\rangle\!\big\rangle \ .
\ee 
As before, we introduce the dimensionless quantities
\be
\label{eq:fluctuations_other_def}
\Sigma^{(\rm{d, \, q })}_N(\tau) \equiv \frac{\sigma_N^{(\rm{d, \, q })}(\tau)}{\Delta_{\mathcal{\hat H}_0}/2} \ .
\ee

It would be desirable to find models of QBs able to reach high values of $\bar E$ and, at the same time,
minimizing the various fluctuations. 
While an overall rescaling of the whole Hamiltonian ($\mathcal{\hat H}\to \alpha \mathcal{\hat H}$) only implies
a redefinition of times, and thus can be easily taken into account, the role played by the relative strength $\kappa$ 
of the charging Hamiltonian is less trivial [see Eq.~\eqref{eq:charging_protocol_general}]. 
A small value of $\kappa$ makes $\mathcal{\hat H}_1$ to be a small perturbation of the global Hamiltonian
$\mathcal{\hat H}$, thus resulting in a low value of $\bar E$. 
In contrast, by increasing $\kappa$, the charging Hamiltonian $\mathcal{\hat H}_1$ becomes a strong perturbation
and may induce transitions from $\ket 0$ to the highly excited states of $\mathcal{\hat H}_0$.

\section{Models}
\label{sec:models}

To unveil the role played by the many-body character of a QB on its charging precision, we have studied a variety
of Hamiltonians, which can be seen as one-dimensional spin-$1/2$ chains (each spin representing a quantum cell,
as discussed above).
These include interactions among the various spins and disorder in the coupling strengths:
as we shall see below, the interplay between two such ingredients is crucial to stabilize the process of energy injection
and thus to any reliable definition of many-body QB.

The first class of charging Hamiltonians is given by
\begin{equation}
  \label{eq:QBham1_MBL}
  \mathcal{\hat H}_1^{\rm (MBL)} = \sum_{j=1}^{N}
  \big( - J_j \, \hat \sigma_j^x \hat \sigma_{j+1}^x + J_2 \, \hat \sigma_j^x \hat \sigma_{j+2}^x \big) \ .
\end{equation}
The first term in the right-hand-side stands for a nearest-neighbor Ising coupling, where the
coefficients $J_j$ are composed of a constant piece plus a static random fluctuation term, $J_j = J + \delta J_j$,
with $\delta J_j$ sampled over a uniform distribution with support in $[- \delta J \, , \, \delta J ]$.
The second term describes a next-to-nearest-neighbor interaction with fixed coupling constant $J_2$.
The Hamiltonian in Eq.~\eqref{eq:QBham1_MBL}, together with the static QB model~\eqref{eq:QBham0},
constitutes a many-body system that exhibits a variety of different quantum phases, ranging from the Anderson localized (AL)
to the many-body localized (MBL), as well as to the ergodic phase. The phase diagram of
$\mathcal{\hat H}_0 + \mathcal{\hat H}_1^{\rm (MBL)}$ has been extensively studied,
since it represents one of the prototypical models of many-body localization-delocalization
transition~\cite{kjall2014, MBL_Nandkishore, MBL_Alet, MBL_Abanin}.



The second class of charging Hamiltonian that we analyze is a highly nonlocal model inspired by the so-called
``Kourkoulou-Maldacena'' SYK model~\cite{kourkoulou2017}.
In its original formulation, the SYK model describes a strongly interacting system of $2N$ Majorana fermions, coupled through
a fully nonlocal, all-to-all, random interaction. 
Denoting with $\hat \gamma_j$ the $j$th Majorana fermion ($\hat \gamma_j = \hat \gamma^\dagger_j$, with $j = 1, \ldots, 2N$),
such that $\{\hat \gamma_i, \hat \gamma_j\} = \delta_{ij}$, the SYK charging Hamiltonian writes
\begin{equation}
  \label{eq:SYK_4_hamiltonian}
  \mathcal{\hat H}_1^{\rm (SYK)} =
  \sum_{i < j < k < l} J_{ijkl} \, \hat \gamma_i \hat \gamma_j \hat \gamma_k \hat \gamma_l \ .
\end{equation}
The couplings $J_{ijkl}$ of the quartic term are randomly Gaussian distributed,
with null mean values and variance
\be
\label{eq:random_couplings_definitions}
\langle\!\langle J_{ijkl}^2 \rangle\!\rangle = \frac{3\,J^2}{4\,N^3} \ .
\ee
The above Hamiltonian can be cast in the spin-$1/2$ setting by employing a Jordan-Wigner transformation (JWT),
\begin{subequations}
\begin{eqnarray}
  \hat \gamma_{2j-1} & = & \frac{1}{\sqrt{2}} \left( \prod_{i=1}^{j-1} \hat \sigma^z_i \right) \hat \sigma^x_j \ , \\
  \hat \gamma_{2j}  & = & \frac{1}{\sqrt{2}} \left( \prod_{i=1}^{j-1} \hat \sigma^z_i \right) \hat \sigma^y_j \ .
\end{eqnarray}
\end{subequations}
Once rewritten in this language, the SYK model looks highly nonlocal and interacting, contrary to the charging
Hamiltonian of the MBL spin chain [Eq.~\eqref{eq:QBham1_MBL}], which instead couples next-to-nearest neighbors, at most.
On the opposite hand, the static ``Kourkoulou-Maldacena'' Hamiltonian $\mathcal{\hat H}_0$, which in the Majorana language reads
\begin{equation}
  \label{eq:local-SYK_majorana}
  \mathcal{\hat H}_{0} = - 2 i \, h \sum_{j \ \mathrm{odd}} \hat \gamma_j \hat \gamma_{j+1} \ ,
\end{equation}
is equivalent to the spin-$1/2$ Hamiltonian of Eq.~\eqref{eq:QBham0}, after using the above JWT.
Each quantum cell of the QB thus corresponds to a couple of neighboring Majorana fermions.

As we shall see in the next sections, these two classes of models exhibit radically different behaviors in terms of QB charging stability.
The reason intimately resides in the nonlocality of SYK, {and on its highly chaotic nature,} with respect to a nearest neighbor or next-to-nearest neighbor spin chain models.

\section{Emergence of fluctuations in the charging process of a QB}
\label{sec:interacting_models}

We now present the results on the different kinds of fluctuations that may emerge during QB charging,
namely temporal, disorder and quantum fluctuations.
To this purpose, we focus on three different models of charging Hamiltonian, which can be obtained either with
Eq.~\eqref{eq:QBham1_MBL} or Eq.~\eqref{eq:SYK_4_hamiltonian}, after suitably tuning the various defining parameters.
For the sake of clarity, in the presentation we will always express energies in units of $h$.
The situations to whom results shown below refer are:
(i) the AL model, corresponding to Eq.~\eqref{eq:QBham1_MBL} with $J_2=0$, $J$, $\delta J = 8.33 \, h$;
(ii) the MBL model~\eqref{eq:QBham1_MBL} with $ J = 1.67 \, h$,  $\delta  J = 8.33 \, h$, and $J_2=0.5 h$;
(iii) the SYK model in Eq.~\eqref{eq:SYK_4_hamiltonian} with $J = h$.

These particular values of the parameters, for the cases (i) and (ii), have been chosen to coincide with the values considered in~\cite{rossini2019}.
However, we have extensively cheched that all the results shown and discussed hereafter are qualitatively the same with respect
to the specific choice of the Hamiltonian parameters, within the same class of model.

\subsection{Temporal fluctuations}

\begin{figure*}[!t]
  \includegraphics[width=5.8cm]{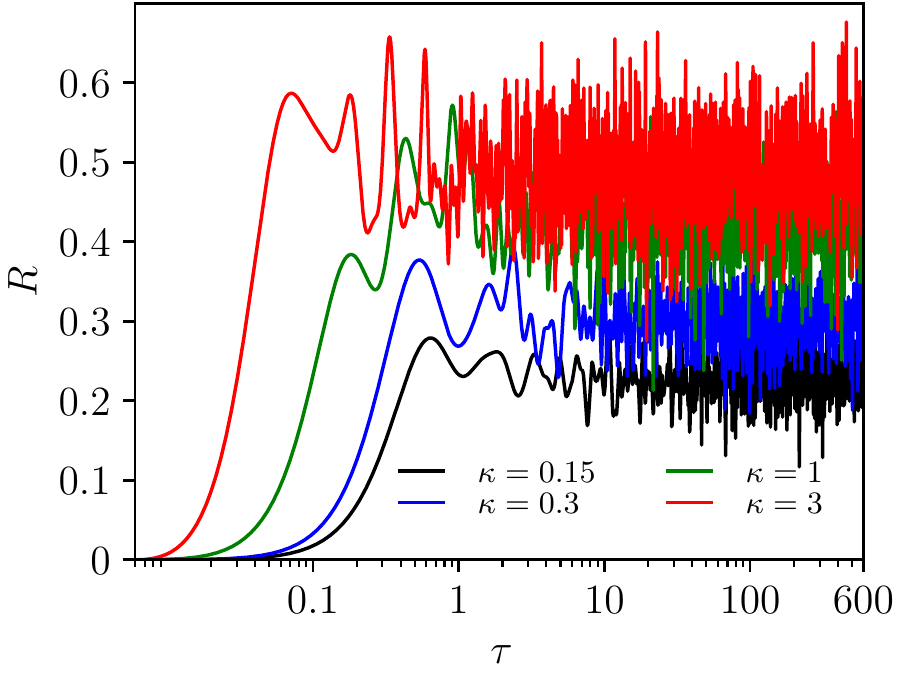}
  \includegraphics[width=5.8cm]{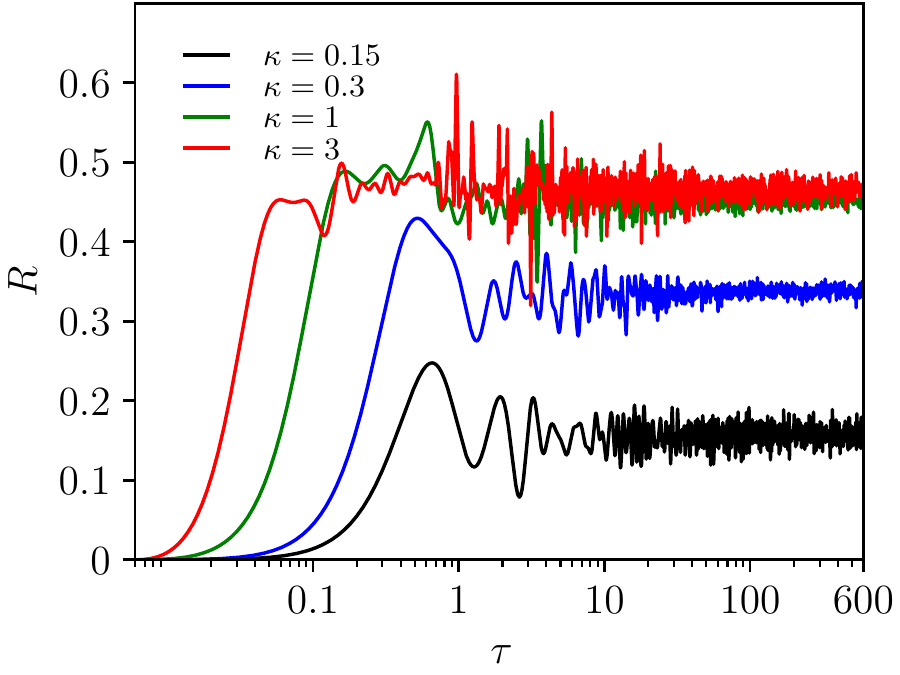}
  \includegraphics[width=5.8cm]{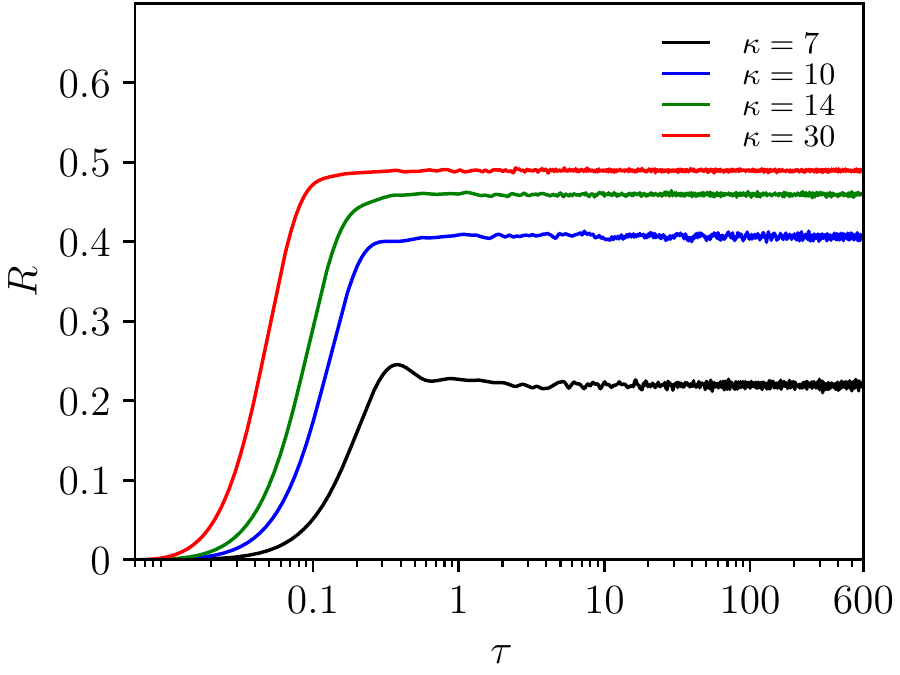}
  \caption{The charging ratio $R(\tau)$ as a function of $\tau$ (measured in units of $h$),
    for a single ensemble realization of the AL spin chain [panel (a)], the MBL spin chain [panel (b)],
    and the SYK Hamiltonian [panel (c)].
    The different curves in the three plots stand for various values of $\kappa$, as indicated in the legends,
    and all are for a QB with $N=15$ quantum cells.}
  \label{fig1}
\end{figure*}

We first focus on the temporal fluctuations of the average energy $E(\tau)$ stored in the QB
through the charging protocol. 
To give a hint on the importance of such kind of noise in the charging performance,
we analyze the time behavior of the ratio $R(\tau)$ between the energy stored in the QB
and the bandwidth of $\mathcal{\hat H}_0$, 
\begin{equation}
  \label{eq:ratio_en_def}
  R(\tau) \equiv \frac{E(\tau)}{\Delta_{\mathcal{\hat H}_0}} \ .
\end{equation}

Figure~\ref{fig1} displays some representative results for the indicator $R(\tau)$ as a function of the charging time $\tau$,
for a single ensemble realization, in the AL, the MBL, and the SYK cases.
The numerical data have been obtained by exact diagonalization and discrete time evolution, using logarithmic time step intervals. We checked that the time step intervals were sufficiently small to ensure the convergence of the results.
The various curves stand for different values of the dimensionless parameter $\kappa$ in the charging
Hamiltonian~\eqref{eq:charging_protocol_general}.
We observe that all the curves grow as a function of time $\tau$ until they saturate to a value corresponding to $\bar{E}$,
whose precise value depends on $\kappa$.
Moreover, by increasing $\kappa$, $\bar E$ increases as well, up to the value $R(\tau) \sim 1/2$,
while any larger $\kappa$ does not help in further increasing $\bar E$, while it simply reduces the value of $\bar \tau$.
This means that, when $\kappa$ is strong enough, the quench Hamiltonian induces a transition from $\ket 0$ to
a superposition involving several eigenstates of $\mathcal{\hat H}_0$, symmetrically distributed around the center
of the bandwidth of $\mathcal{\hat H}_0$, thus ensuring that $\bar E \sim \Delta_{\mathcal{\hat H}_0}/2$.
On the other hand, the temporal fluctuations are mostly unaffected by $\kappa$ and one has to find smarter ways to reduce them. 
This set of fluctuations will be the main focus of the analysis, and we will show how they can be efficiently suppressed.
As we will see, the internal structure of $\mathcal{\hat H}_1$ will play a crucial role for this task.

In Ref.~\cite{rossini2019}, it was argued that the presence of interactions in $\mathcal{\hat H}_1$
can help in reducing the temporal fluctuations during the charging of the QB.
Here we show that, while local interactions have only limited effects on the fluctuations,
nonlocal correlations allow to build models of QBs with high temporal stability in $\bar E$.
These qualitative features are already visible from Fig.\ref{fig1}, where we clearly see that by increasing the non locality degree of the interactions (from left to right panel) temporal fluctuations are greatly reduced. 
One of the main goals of this paper will be to explain such behavior and to quantitatively describe it.

We now fix the value of $\kappa$ and first comment on the optimal charging time of the QB,
$\langle\!\langle \bar \tau \rangle\!\rangle$, evaluated as the time at which the energy stored in the QB
reaches a value equal to $99 \%$ of the average energy
(we have tested that the results are not affected by the arbitrary choice of this cutoff),
as well as the corresponding energy $\langle\!\langle \bar E \rangle\!\rangle$.
We now refer to quantities averaged over several realizations of $\mathcal{\hat H}_1$.
Our numerical simulations indicate (not shown) that, {for all the three cases considered},
the optimal charging time is a decreasing function with the number $N$ of cells,
while the averaged energy stored in the battery at the optimal time scales linearly with the number of sites.
The last property easily follows from the fact that, as already pointed out in Sec.~\ref{sec:charging_generalities},
when $\kappa$ is large enough, $\langle\!\langle \bar E \rangle\!\rangle$
is determined by the bandwidth of $\mathcal{\hat H}_0$, which scales linearly with $N$.
These two observations agree with the results obtained in Ref.~\cite{rossini2019} for the MBL model,
and certify that all our models are indeed able to properly charge the battery.

\begin{figure*}[ht]
  \centering
  \subfigure[\label{fig:fig3_MBL}]{\includegraphics[width=8cm]{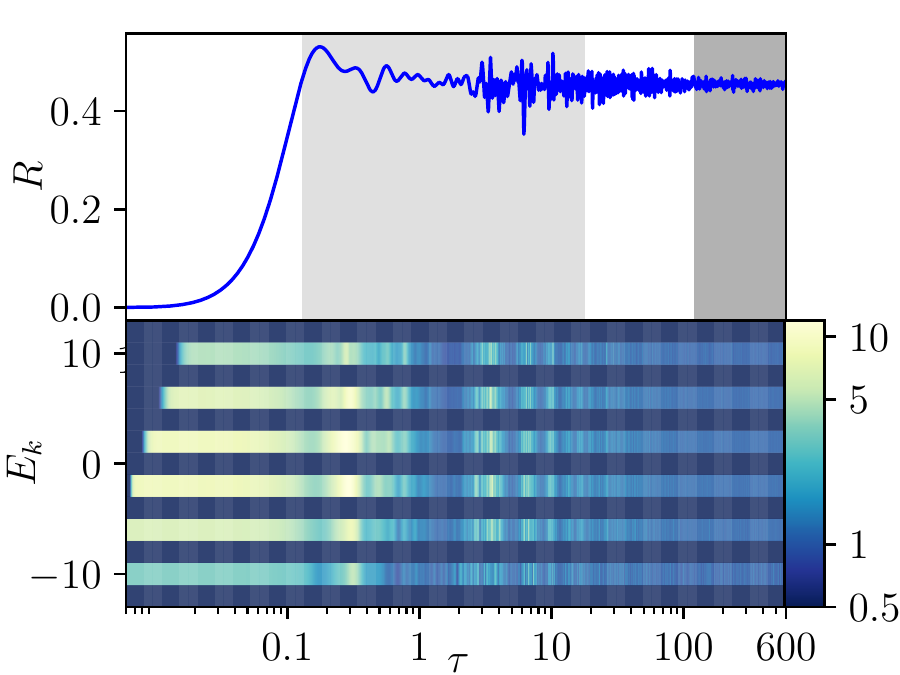}}
  \subfigure[\label{fig:fig3_l-SYK}]{\includegraphics[width=8cm]{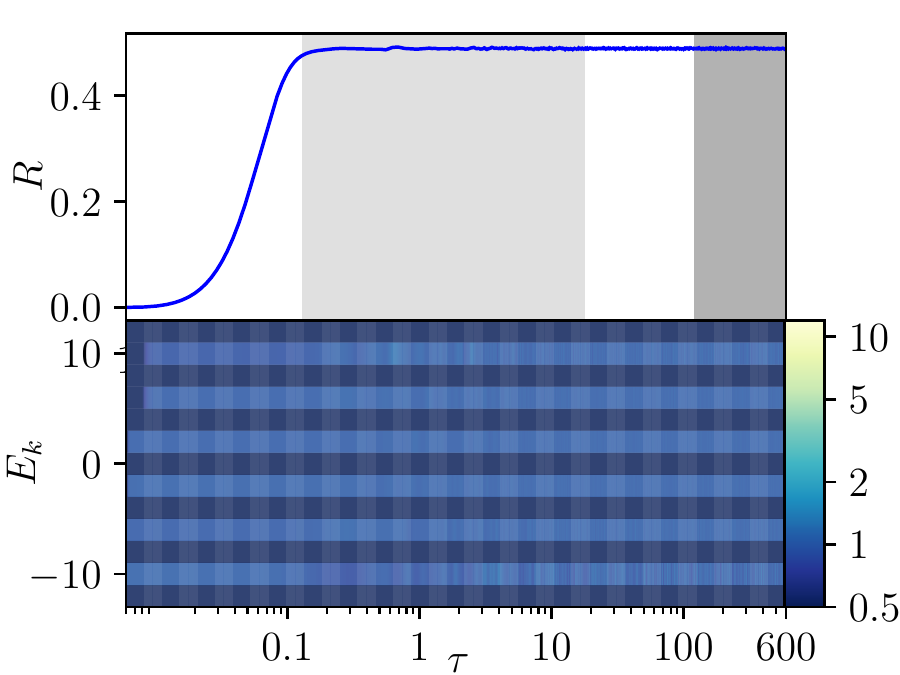}}\\
  \subfigure[\label{fig:MBL_fluctuations}]{\includegraphics[width=8cm]{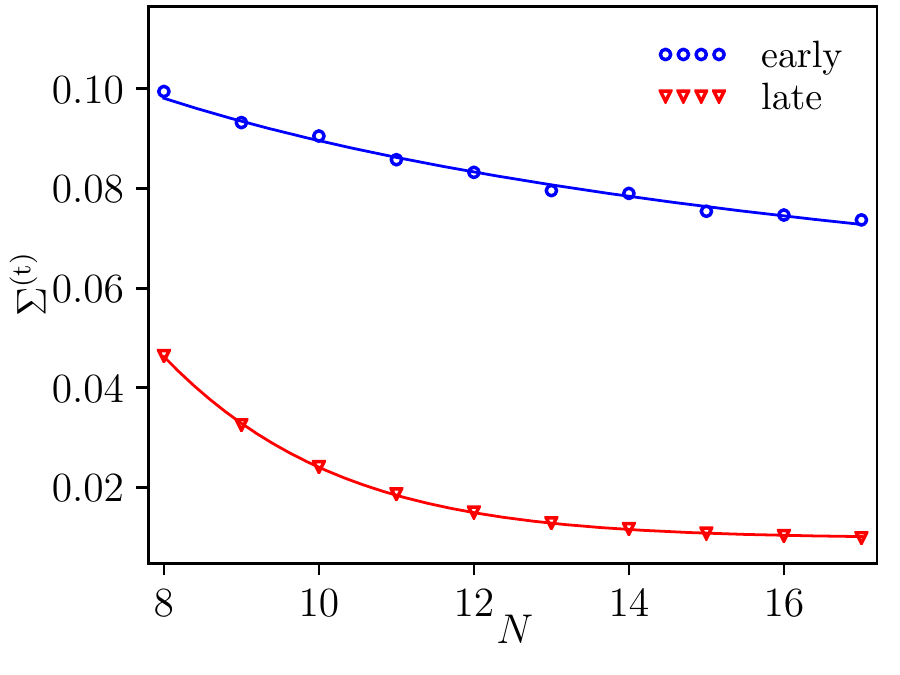}}
  \subfigure[\label{fig:l-SYK_fluctuations}]{\includegraphics[width=8cm]{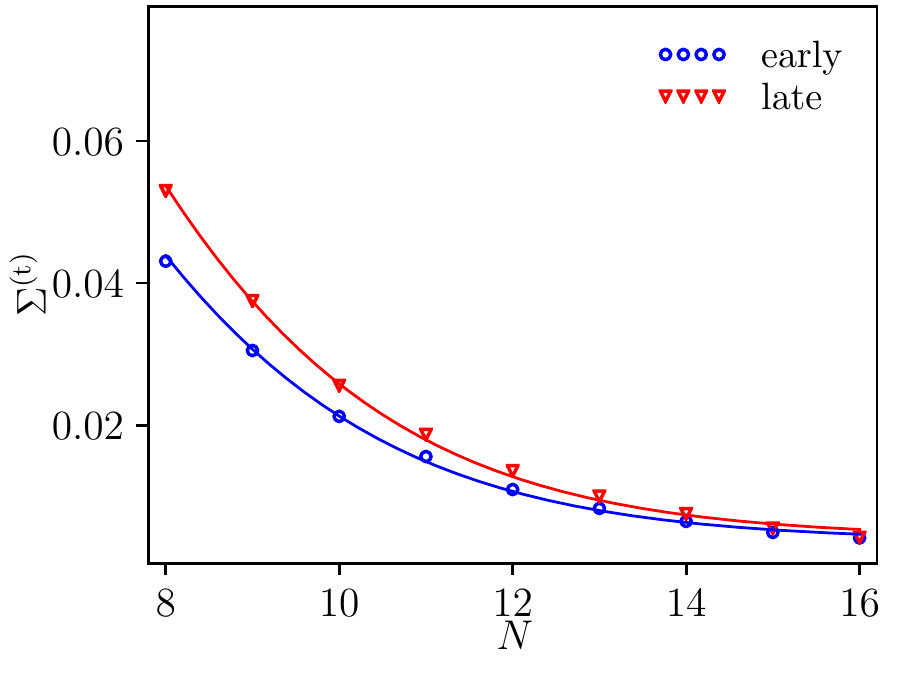}}
  \caption{Analysis of the temporal fluctuations during the QB charging protocol through the MBL model with $N=15$ spins
    (left panels) and the SYK model with $2N = 30$ Majorana fermions (right panels).
    Panels~\subref{fig:fig3_MBL} and~\subref{fig:fig3_l-SYK}:
    The functions $R(\tau)$ (upper part) and $\sigma_k$ in the various energy sectors (lower part) as a function of time,
    for a single realization of the random variables.
    Shaded areas denote the time intervals for the early- and late-time window
    (light grey and dark grey, respectively) analyzed below --- see text.
    Here $\bar \tau = 0.14$ and $0.12$, in the two panels.
    Panels~\subref{fig:MBL_fluctuations} and~\subref{fig:l-SYK_fluctuations}:
    The early- and late-time temporal fluctuations, $\Sigma_N^{(\rm t)} (\tau_1,\tau_2)$ of Eq.~\eqref{eq:fluctuations_temp_def},
    for the MBL and the SYK models, as a function of the number $N$ of QB cells.
    Continuous lines correspond to the fits~\eqref{eq:early_fluct_MBL}-\eqref{eq:late_fluct_MBL}
    and~\eqref{eq:early_late_fluct_l-SYK}, respectively.
    Results in the two bottom panels have been obtained by averaging over $500$~\subref{fig:MBL_fluctuations}
    and $100$~\subref{fig:l-SYK_fluctuations} ensemble realizations.}
    \label{fig3}
\end{figure*}

To obtain a more quantitative assessment of the role of temporal fluctuations in the QB charging mechanism,
we have performed further extensive simulations for the MBL and the SYK model, by fixing the constant $\kappa$.
The corresponding analysis of the AL model is not reported in the main text, since already from Fig.~\ref{fig1} is clear that it shows huge temporal fluctuations at all times. For sake of completeness, this is reported in App.~\ref{app:And_time}.

In order to make fair comparisons between these models, we have set the constant $\kappa$ in the MBL QB equal to one,
while for the SYK QB it has been fixed in such a way that the two quench Hamiltonians have the same bandwidth,
$\Delta_{\mathcal{\hat H}_1^{\rm {MBL}}} = \Delta_{\mathcal{\hat H}_1^{(\rm SYK)}}$.
The results of our analysis are reported in Fig.~\ref{fig3}.

The upper part of panels~\ref{fig:fig3_MBL} and~\ref{fig:fig3_l-SYK} display the ratio $R(\tau)$ of
Eq.~\eqref{eq:ratio_en_def}, for a single ensemble realization of the MBL spin chain and the SYK battery, respectively.
We immediately recognize that, compared to the analogous plot for the Anderson spin chain [c.f.~Fig.~\ref{fig1}(a), the green curve],
the MBL battery is able to partially reduce the temporal fluctuations at late times,
{\it i.e.}~for times much larger than the optimal charging time $\tau \gg  \, \langle\!\langle \bar \tau  \rangle\!\rangle $.
However, at early times, {\it i.e.} for times roughly included in the light grey areas
in the panels,
\begin{equation}
  1 \lesssim \frac{\tau}{\langle\!\langle \bar \tau \rangle\!\rangle} \lesssim 20 \ , \qquad
  \mbox{(early-time window)},
\end{equation}
fluctuations are still very large. On the contrary, the plot clearly shows that the SYK battery
is extremely precise and stable at any time scale: all the temporal fluctuations,
after reaching $\langle\!\langle \bar \tau  \rangle\!\rangle$, are completely removed.
To analyze this fact we shall also define the following time interval,
identified by the dark grey areas in the panels,
\begin{equation}
  120 \lesssim \frac{\tau}{\langle\!\langle \bar \tau \rangle\!\rangle} \lesssim 600 \ , \qquad
  \mbox{(late-time window)}.
\end{equation}
It should be stressed that the early-time window is very relevant for energy storage purposes, since one would like
to have a great control of the charging precision immediately after reaching the saturation of the energy stored.
We stress that the precise values of the early and late time windows are not important, and the details of these choices do not affect qualitatively the behaviors we are discussing.

The two plots~\ref{fig:fig3_MBL} and~\ref{fig:fig3_l-SYK} clearly unveil the qualitative advantage of the SYK model,
and confirm the intuition that nonlocal correlations play a crucial role in the charging dynamics.
Thus, a strongly interacting, nonlocal, quench (like the SYK model) represents a perfect candidate to build models
of very stable QBs with high charging precision.
Moreover, in {Sec.~\ref{sec:quantum_chaos}} we show that nonlocality alone in $\mathcal{\hat H}_1$ is not enough,
and that the highly chaotic dynamics of the SYK system is crucial in order to efficiently suppress the temporal fluctuations.
In Appendix~\ref{app:l-SYKvsml-SYK} we analyze the role of the static Hamiltonian $\mathcal{\hat{H}}_0$,
by making a comparison with another kind of SYK-like QB with a nonlocal $\mathcal{\hat H}_0$
and showing that the two models are qualitatively analogous.
Therefore the precise form of $\hat{ \mathcal{H}}_0$ does not play a major role in the charging dynamics.

\subsubsection{Fluctuations in terms of transition amplitudes}
\label{sec:microscopic}

Let us now have a closer inspection at the microscopic origin of the improved efficiency and charging stability,
in the presence of nonlocal correlations.
In general, temporal energy fluctuations are caused by transitions of the probability amplitudes
\begin{equation}
  c_{k , i} \equiv \langle k , i \, | \, \psi (t) \rangle \ ,
\end{equation}
between eigenstates with different energies.
Here $\{ \ket{k , i} \}$ denote the eigenstates of $\mathcal{\hat H}_0$; the index $k$ labels the energy level
$E_k$, while $i=1, \ldots, d_k$ counts the different degenerate eigenstates within the same eigenspace ($d_k$ is the
corresponding degeneracy degree).
Such transitions cause a large fluctuation of the energy stored if both the following conditions are met:
(i) The eigenstates have very different energies;
and (ii) The probability amplitudes of being in the eigenstates involved in the transition are large.

The first condition is immediate to understand: if two eigenstates have similar energies, the energy stored
in the battery will not vary much after the transition (the extreme case would be a transition between degenerate eigenstates).

The second condition is more subtle: let us consider the limit case in which the evolved ket, $\ket{\psi(t)}$,
can be written as a superposition of all the eigenstates of $\mathcal{\hat H}_0$
with approximately the same probability amplitudes
\be
\label{eq:ci_complete_superposition}
c_{k , i} \sim \frac{1}{\sqrt{D}} \ ,
\ee
where $D$ is the dimension of the system's Hilbert space.
Since $D=2^N$ is exponentially large in the number $N$ of cells,
all the coefficients $c_{k , i}$ will be very small.
In this case, a transition between eigenstates, even with very different energies, will not be reflected in large fluctuations.
Indeed, since the bandwidth of $\mathcal{\hat H}_0$ scales linearly in $N$, the fluctuation 
\be
\label{eq:energy_fluctuation_small}
\Delta E(t) \sim c_{k,i}^2 N \sim O\bigr(2^{-N}\bigr) \ ,
\ee
will be, at most, exponentially small in $N$.
In contrast, in the situation where only just few (of order $N$) eigenstates of $\mathcal{\hat H}_0$
are involved in the expansion of the evolved state, some of the coefficients $c_{k, i}$ can be relatively large
\be
c_{k , i} \sim \frac{1}{\sqrt{N}} \ ,  \ \mathrm{for \ some} \ k, \, i \ ,
\ee
and a  transition including one of these states will cause a large fluctuation in the energy stored.

Given these considerations, we expect that in the MBL case the evolved state at early times should have non vanishing overlap with just few eigenstates of $ \mathcal{\hat H}_0$ (for each energy level), while involving more and more states at late times, thus reducing the associated fluctuations.
On the contrary, for the SYK model the evolved state should involve a large portion of the Hilbert space of $\mathcal{\hat H}_0$ from the very early times.

To corroborate this hypothesis, we first notice that the energy spectrum of $\mathcal{\hat H}_0$
is formed by several lines, well separated from each other. 
Each line having energy $E_k$ has a degeneracy degree $d_k$ counted
by the number of configurations with the right number of aligned spins.
Hence, to estimate if, for a given energy eigenvalue, the evolved state has non vanishing overlap
with just few or many eigenstates of $\mathcal{\hat H}_0$, we consider the quantity
\begin{equation}
  a_k^i(\tau) \equiv | \braket{k , i \, | \, \psi(\tau)} |^2 \ ,
  \label{eq:overlap}
\end{equation}
which expresses the probability of measuring the evolved state in one of the eigenstates $\ket{k,i}$
associated to the level $E_k$. 
We have taken into account all of such eigenstates, {\it i.e.}~with fixed $k$
and varying $i=1, \ldots , d_k$, and computed the standard deviation associated to the
corresponding quantities~\eqref{eq:overlap}, divided by their average value.
Namely,
\begin{equation}
  \sigma_k(\tau) = \frac{1}{\overline{a_k(\tau)}} \, \sqrt{ \frac{1}{d_k - 1} \sum_{i = 1}^{d_k} \,
    \Big[ a_k^i(\tau) - \overline{a_k(\tau)} \Big]^2 } \ ,
\end{equation}
with $\overline{a_k(\tau)} = \big[ \sum_{i = 1}^{d_k} a_k^i(\tau) \big] / d_k$.

We can thus determine if the expansion of $\ket{\psi(\tau)}$ in the degenerate eigenstates for a given energy level
is involving many or few of the eigenstates, with the former case corresponding to small values $\sigma_k$
and the latter associated to large values of $\sigma_k$. 
The results are reported in the lower parts of panels~\ref{fig:fig3_MBL} and~\ref{fig:fig3_l-SYK},
for the MBL and the SYK model, and confirm our conjecture:
the MBL system shows at early times huge values of $\sigma_k$ for each energy sector, and in correspondence with these peaks we can clearly trace a huge temporal fluctuation of the average energy stored in the battery (see the upper panel of the figure).
This behavior gets reduced by increasing time and, after bouncing for a while, the system reaches low values for all the $\sigma_k$s.
On the other hand, from the very beginning, the SYK model displays low values for $\sigma_k$ (around one order of magnitude smaller), clearly showing that in this model many more eigenstates of $\mathcal{\hat H}_0$, for each energy level, are rapidly involved in the expansion of the evolved state.
Hence, the charging protocol  turns out to be very stable, and this is reflected in the very small temporal fluctuations. 
It should be emphasized that the low values of all the $\sigma_k$s, for the SYK model, are reached at a time scale which is even shorter than the optimal charging time, $\langle\!\langle \bar \tau  \rangle\!\rangle$, thus ensuring the total absence of temporal fluctuations.

This microscopic argument confirms that temporal fluctuations get suppressed when an initially localized state (in the eigenbasis of $\mathcal{\hat H}_0$) spreads and covers a very large portion of the eigenstates of $\mathcal{\hat H}_0$ and, as such,  we think that it could be naturally linked to the physics of scrambling and of thermalization.
Indeed, the thermalization properties of the SYK model have been already investigated in Refs.~\cite{eberlein2017,bhattacharya2018}, where it has been demonstrated that this model shows thermalization, even without long time averaging. This fact corresponds to a quantum version of mixing, a much stronger phenomenon as compared to ergodicity~\cite{bhattacharya2018}.
On the other hand, a MBL system does \emph{not} thermalize in the thermodynamic limit. 
Hence, we expect that the huge suppression of the temporal fluctuations at late times, in this case, should be a finite $N$ effect: by increasing the size of the system, the time at which the fluctuations are highly suppressed should tend to infinity.
This is consistent with the results, reported in Fig. 4(b), of Ref.~\cite{rossini2019}, where an increase in the  temporal fluctuations of the MBL model (without separating early and late times) could be observed at the largest values of $N$.

\subsubsection{Temporal fluctuations of the charging energy in the early- and late-time windows}
\label{sec:fluctuations_fits}

So far we have argued that, in general, during the charging process of a generic QB, two different time windows
can be identified: after reaching the optimal charging time, $\langle\!\langle \bar \tau \rangle \!\rangle$,
we have an ``early-time'' window, in which the averaged energy stored in the battery, $E (\tau)$,
undergoes huge temporal fluctuations, the expansion of the evolved state $\ket{\psi(\tau)}$
on the basis of the eigenstates of $\mathcal{\hat H}_0$ involves just few eigenstates for each energy level. 
On much larger time scales, the dynamics turns to a ``late-time'' window, in which the energy $E (\tau)$
displays suppressed temporal fluctuations, the evolved state has spread to cover a large portion
of the eigenstates of $\mathcal{\hat H}_0$.
We have also argued that the time of crossover, between the early time and the late time behavior, is connected with the thermalization properties of the system under investigation and, as such, it is model-dependent. 

We now turn to a more explicit evaluation of the temporal fluctuations in the charging energy,
[cf. Eq.~\eqref{eq:temporal_variance}], in the two time windows defined before.
To be precise, we address the dimensionless quantity $\Sigma_N^{(\rm t)} (\tau_1,\tau_2)$
of Eq.~\eqref{eq:fluctuations_temp_def} in the two time windows (early and late) over which we take the time integral.
In Fig.~\ref{fig:MBL_fluctuations} we plot the results for the MBL spin chain.
The behavior for different $N$, of \eqref{eq:fluctuations_temp_def}, at early and late times is qualitatively different: while at late times $\Sigma$ is fastly decreasing with $N$, at early time $\Sigma$ instead shows a much slower decrease.
In Fig.~\ref{fig:l-SYK_fluctuations} we plot the results for the SYK battery. 
Here, the situation is  different: both the early and late time fluctuations are rapidly suppressed in $N$ 

These observations can be made quantitative: in the MBL case, the early time curve is greatly reproduced by the function
\be
\label{eq:early_fluct_MBL}
\Sigma^{(\rm t)}_N = \frac{a}{\sqrt N} + b \ ,
\ee
with $a$ and $b$ being fitting parameters. 
On the other hand, the late time behavior is well reproduced by
\be
\label{eq:late_fluct_MBL}
\Sigma^{(\rm t)}_N = a \, N^2 \, 2^{- N} + b \ ,
\ee
which shows that the late time temporal fluctuations are exponentially suppressed with $N$.

The numerical data for the SYK case instead can be reproduced by the function
\be
\label{eq:early_late_fluct_l-SYK}
\Sigma^{(\rm t)}_N = a \, N^{2.5} \, 2^{- N} + b \ .
\ee
In summary, this shows that the temporal fluctuations, in the SYK model, \emph{both} at early and late times,
are exponentially suppressed by increasing the size of the battery.
On the other hand, the MBL battery, shows this exponential suppression only at late times, while at early times
it follows a $1/\sqrt N$ suppression factor, only.
The exponential suppression, at early times, of SYK battery, makes clear that with this model it is possible to obtain very stable charging protocols, in which the average energy stored in the battery is essentially determined with very high precision, even with relatively small batteries.

It has been shown in \cite{llobet2019} (in the similar context of work extraction) that precisely an exponential and a $1/\sqrt N$ suppression factors are associated to, respectively, collective processes, {\it i.e.} processes in which all the cells are collectively controlled in the protocol, and single cells protocols, in which each cell is individually processed.
It is then natural to expect that, due to the nonlocal nature of its hamiltonian, the SYK has a genuine collective dynamics from the very early times, shorter than the optimal charging time $\langle\!\langle \bar \tau \rangle\! \rangle$, while the MBL battery needs a certain amount of time to start a collective dynamics, with an initial single-body behavior.
{On this respect, we conclude that an integrable Hamiltonian, like the AL spin chain in the left panel of Fig.~\ref{fig1}, never reaches a collective dynamics.
In Sec.~\ref{sec:quantum_chaos} we will confirm this intuition, by showing that the presence of quantum chaos in the quench Hamiltonian is necessary to ensure the exponential suppression of the fluctuations.} 

This is in perfect agreement with the microscopic description of the fluctuations we have provided in Sec.~\ref{sec:microscopic}, and with the findinds of Sec.~\ref{sec:ergotropy}, where we show that the MBL battery needs a large amount of time in order to involve a large portion of the eigenstates of $\mathcal{\hat H}_0$ in the expansion of the evolved state $\ket{\psi(\tau)}$.
Moreover, the absence of a crossover in the SYK system can be again understood in terms of the microscopic description of Sec.~\ref{sec:interacting_models}, where we observed that for the SYK system $\ket{\psi(\tau)}$ involves a large portion of the Hilbert space from times which are smaller than the optimal charging time, thus ensuring that \emph{all} the temporal fluctuations are exponentially suppressed.

\subsection{Disorder and quantum fluctuations}

\begin{figure}[!t]
  \subfigure[\label{fig:dis_early}]{\includegraphics[width=8cm]{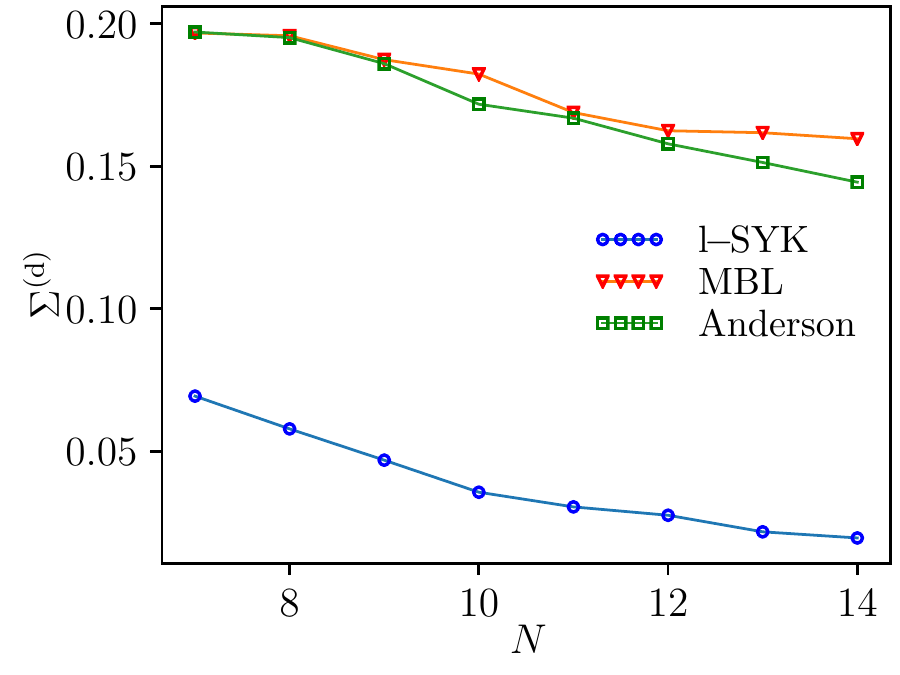}}
  \subfigure[\label{fig:dis_late}]{\includegraphics[width=8cm]{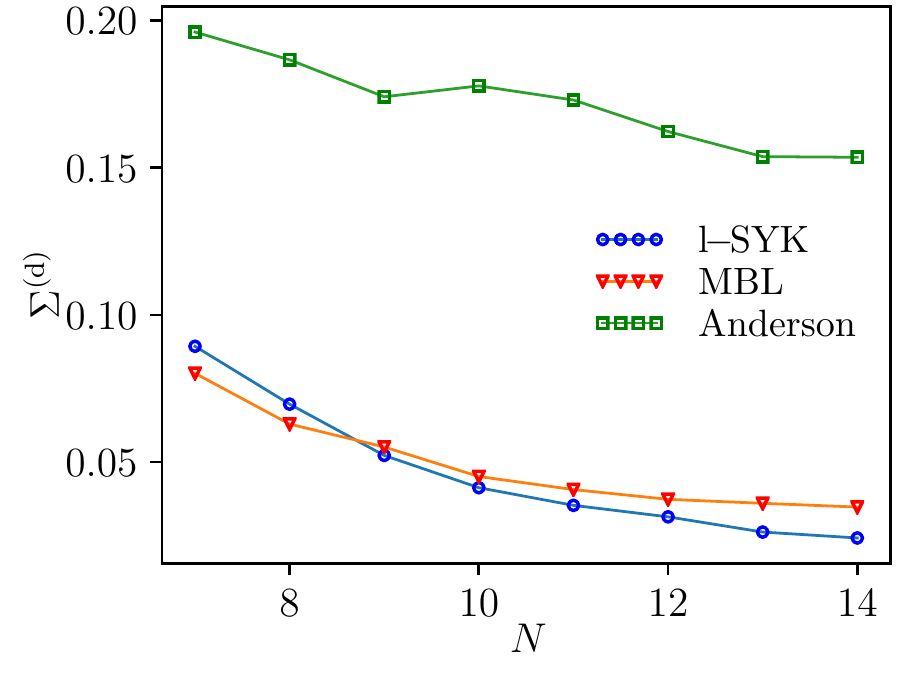}}
  \caption{\subref{fig:dis_early}: The disorder fluctuations at early times, as measured by $\Sigma^{(\rm d)}$ and as functions of the lattice size. 
    \subref{fig:dis_late}: Same quantity for for the late time window.
    The results are obtained by averaging over $500$ (up to $N =11$), $250$ ($N = 12 , \, 13$) and $100$ ($N = 14$) ensemble realizations.
    We have chosen the following values of the time, $\tau_1 \equiv 1.5 \, \bar \tau$ for the early time window and $\tau_2 \equiv 0.5 \, \bar \tau_{\mathrm{max}}$ for the late time window, with $\tau_{\mathrm{max}}$ the final time, $\tau = 600$.
  }
  \label{fig:figdisorder}
\end{figure}

We now move to the discussion of the disorder and quantum fluctuations for all the three models considered.
From~\eqref{eq:disorder_variance} and~\eqref{eq:quantum_variance}, we see that both these quantities have to be evaluated at a definite time, $\tau$.
Hence, we have chosen two fixed values of the time, one in the early time window and one in the late time window, to evaluate them.

The results for the disorder fluctuations $\Sigma_N^{(\rm{d})}(\tau)$, cf. Eq.~\eqref{eq:fluctuations_other_def},
are reported in Fig.~\ref{fig:figdisorder}.
In agreement with the previous results, disorder fluctuations for the SYK model are always small,
both at early times and at late times. 
Similarly, by considering the Anderson model, we see that the disorder fluctuations are always large, both at early and late time.
Finally, the MBL system shows a crossover when passing from the early time window to the late time window: it shows a behavior similar to the Anderson spin chain at early times and it moves to a behavior very similar to the SYK model at late times.

\begin{figure}[!t]
  \subfigure[\label{fig:quant_early}]{\includegraphics[width=8cm]{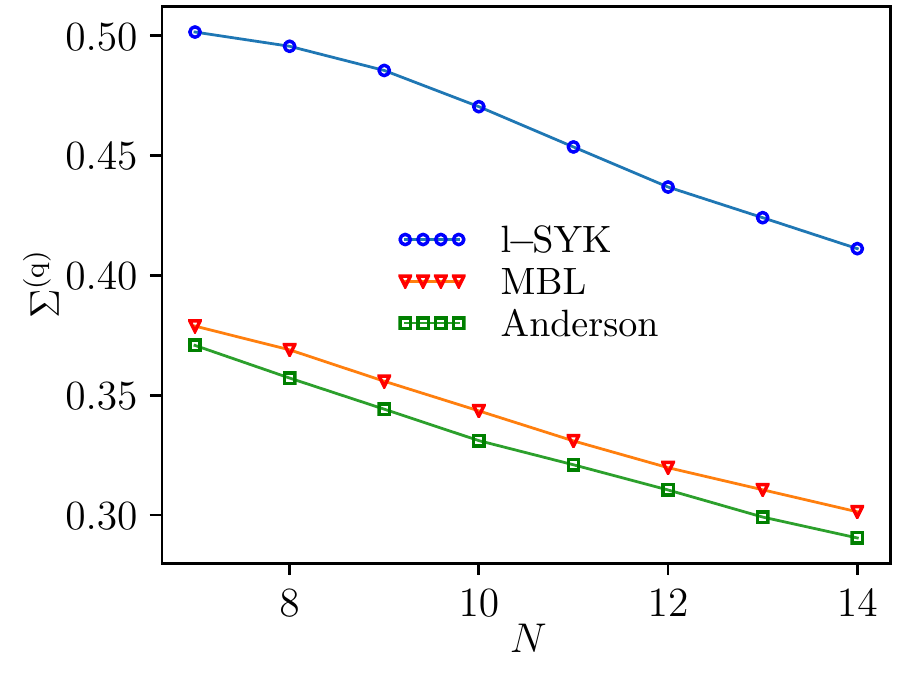}}
  \subfigure[\label{fig:quant_late}]{\includegraphics[width=8cm]{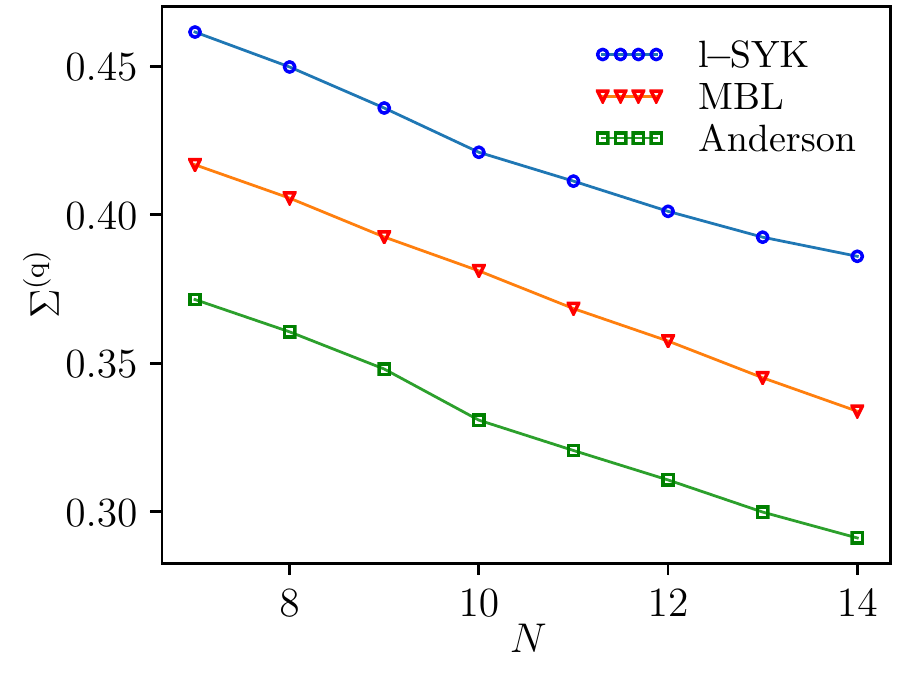}}
  \caption{\subref{fig:quant_early}: The quantum fluctuations at early times, as measured by $\Sigma^{(\rm q)}$ and as functions of the lattice size. 
    \subref{fig:quant_late}: Same quantity for for the late time window.
    The results are obtained by averaging over $500$ (up to $N =11$), $250$ ($N = 12 , \, 13$) and $100$ ($N = 14$) ensemble realizations.We have chosen the following values of the time, $\tau_1 \equiv 1.5 \, \bar \tau$ for the early time window and $\tau_2 \equiv 0.5 \, \bar \tau_{\mathrm{max}}$ for the late time window, with $\tau_{\mathrm{max}}$ the final time, $\tau = 600$.
  }
  \label{fig:figquant}
\end{figure}

Moving to quantum fluctuations, $\Sigma_N^{(\rm{q})}(\tau)$, cf. Eq.~\eqref{eq:fluctuations_other_def},
the results are reported in Fig.~\ref{fig:figquant}.
Again, we find that the MBL model shows a crossover when moving from the early time window to the late time window, becoming more similar to the SYK behavior only at late times.
We also see that the quantum fluctuations for the SYK model are larger than for all the other models we have considered.
Furthermore, it is worth to notice that for all the three models, the values of the temporal and disorder fluctuations are quite similar, while the quantum fluctuations are always larger than the other sources of fluctuations, reaching the largest value of $ \sim 0.50$ for the SYK battery.
This large value of quantum fluctuations for the SYK model could be put in relation with its high charging power, \cite{rossini2019b}, since it has been recently observed in Ref.~\cite{garcia-pintos2019} that high levels of quantum fluctuations are necessary in order to increase the charging power of a QB.

For both the disorder and the quantum fluctuations, we see that they are suppressed by increasing the size of the battery.

{
\section{The role of quantum chaos on the charging stability}
\label{sec:quantum_chaos}
We now elucidate the role that quantum chaos plays in the suppression of the temporal fluctuations, \textit{i.e.} in the charging stability of a QB.

To this end, it is instructive to consider two slightly different models of QBs.
In both cases, the unitary charging protocol is given by:
\be
\label{eq:quadratic_protocol}
\mathcal{\hat H}_{\mathrm{SYK}_2} \equiv  \mathcal{\hat H}_0 + \kappa \, \lambda(t) \, \mathcal{\hat H}_2^{s} \ ,
\ee
where $\mathcal{\hat H}_0$ is the local constant Hamiltonian \eqref{eq:QBham0}.
The quench terms $\mathcal{\hat H}_2^{s}$ with $s= \mathrm{F, B}$, instead, are random mass Hamiltonians, that means quadratic in the field operators, the two differing for their fermionic/bosonic statistics (see below). The case $\mathcal{\hat{H}}_2^{\mathrm{F}}$  is defined by
\be
\label{eq:SYK_2_hamiltonian}
\mathcal{\hat H}_2^{\mathrm F} = i \, \sum_{i < j} K_{ij} \, \hat \gamma_i \hat \gamma_j \ ,
\ee
with the random couplings $K_{ij}$ having null mean values and variances
\be
\label{eq:random_couplings_definitions}
\langle\!\langle K_{ij}^2 \rangle\!\rangle = \frac {1}{2N} \  .
\ee 
On the other hand, $\mathcal{\hat H}_2^{\mathrm{B}}$ is built using the following, \textit{real}, hard-core bosonic operators, $\hat{\chi_i}$, with $i  = 1, \dots , 2N $, which satisfy the following algebra:
\begin{align}
  \label{eq:bosons_algebra}
  \{ \hat \chi_{a} , \, \hat \chi_{b} \} &=
  \begin{cases}
      0 & \text{if }(a, b)=(2i-1, 2i) \text{ for } 1 \leq i\leq N\\
      1 & \text{if }a=b
  \end{cases} \nonumber\\
  \big[ \hat \chi_a , \, \hat \chi_b \big]  &=
  \begin{cases}0 & \mathrm{otherwise}. \end{cases}
  \end{align}
The bosonic quench Hamiltonian is then given by
\begin{equation}
\label{eq:SYK_2_hamiltonian_bosonic}
\mathcal{\hat H}_{2}^\mathrm{B} =  \sum_{i < j} (i)^{s(i , j)} \,  K_{ij} \, \hat \chi_i \hat \chi_j \ ,
\end{equation}
with the same random coupling constants $K_{ij}$ as in \eqref{eq:random_couplings_definitions}.
The factor $s(i , j)$ reads
\begin{equation}
    \label{eq:sign_function_bosons}
s(i , j) \equiv   (1 + (-1)^{j}) \, \delta_{i + 1 ,\,  j} \ ,
\end{equation}
and it ensures that $ \mathcal{\hat H}_{2}^\mathrm{B}$ is Hermitian.

Although very similar, being both quadratic in the field operators, the two quench Hamiltonians \eqref{eq:SYK_2_hamiltonian} and \eqref{eq:SYK_2_hamiltonian_bosonic} have very different properties: the fermionic system is indeed integrable, while the bosonic model is chaotic.
This can be verified by inspecting simple quantum chaos diagnostics, like the so-called r-statistics \cite{Oganesyan_2007}, as we show in App.~\ref{subsec:quantum_chaos_SYK}.

This difference has a direct consequence on the QB performance, as shown in Fig.~\ref{fig:SYK_2_bosonic_fermionic_single_charge} where the charging for a single realization of the coupling constants $K_{ij}$ is reported.
\begin{figure}[!t]
  \includegraphics[width=8cm]{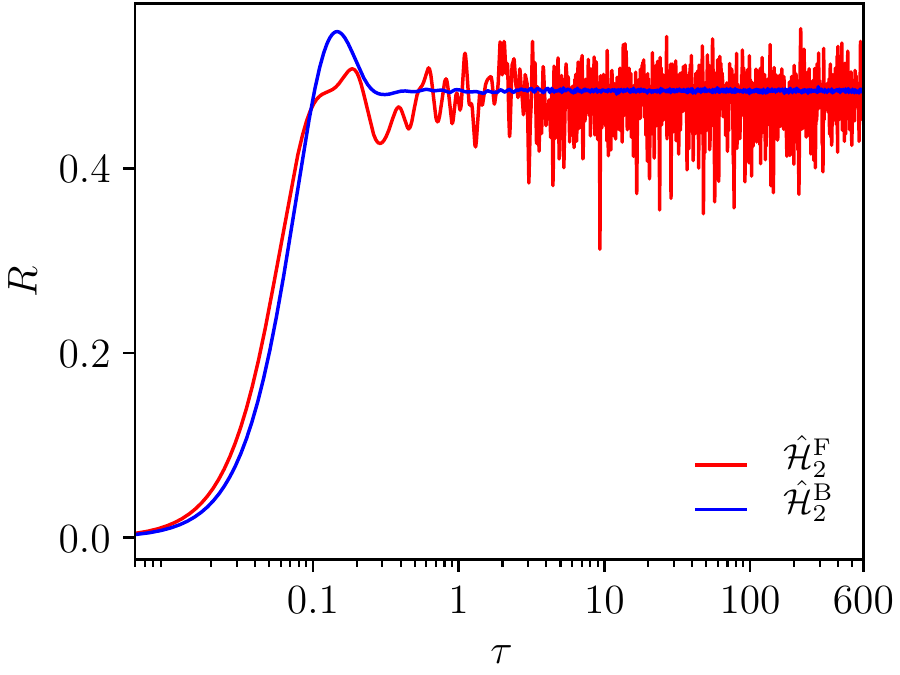}
   \caption{ The charging ratio $R(\tau)$ —see Eq.~\eqref{eq:ratio_en_def}— as a function of $\tau$, for a QB described by the Hamiltonian in Eq.~\eqref{eq:quadratic_protocol}. We study a single realization of the coupling constants $K_{ij}$, for both the bosonic quadratic model, Eq.~\eqref{eq:SYK_2_hamiltonian_bosonic}, and the fermionic model, Eq.~\eqref{eq:SYK_2_hamiltonian}, for $N = 15$.}
  \label{fig:SYK_2_bosonic_fermionic_single_charge}
\end{figure}
As usual, we have fixed the constant $\kappa$ to ensure that $ \mathcal{\hat H}_{2}^\mathrm{B}$ and $ \mathcal{\hat H}_{2}^\mathrm{F}$ have the same bandwidth.
From the figure, one can clearly see that, despite the striking similarity of the two models, the temporal stabilities are completely different, with the bosonic model, being chaotic, which efficiently truncates the fluctuations while the fermionic one shows large fluctuations even at late time.
This shows that quantum chaos is needed to reach charging stability.
Given this result, a natural question which arises is whether more conventional chaotic models, like spin chain models in the ergodic phase, show the same level of charging stability of the SYK quench defined in \eqref{eq:SYK_4_hamiltonian}.
As shown in App.~\ref{subsec:ergodic_vs_SYK}, however, it turns out that the charging stability of the SYK QB is exceptional and definitely larger than the level of charging stability reached by an ergodic spin chain.

It thus remains to understand which particular feature of the chaotic SYK Hamiltonian is responsible for the extremely efficient suppression of the temporal fluctuations.
While a detailed understanding of this point is beyond the scope of the paper, we observe here that the short-range chaos observables, like the r-statistics, simply tell us whether a certain Hamiltonian, at very small energy scales (or equivalently for very long times), shows the same spectral correlations as predicted by random matrix theory (RMT).
However, they say nothing about how large in energy, or equivalently how short in time, the agreement with RMT persists.

To address this issue, which in a sense defines how strong is the chaotic nature of a system, one has to study the so-called long range chaos diagnostics, like the spectral rigidity or the associated spectral form factor \cite{berry1985}.
On this respect, in \cite{gharibyan2019} and \cite{Jia:2019orl}, it has been observed that the Thouless time, which is the time scale at which the agreement with RMT becomes manifest, for the SYK model is very small, of order $\log N$, while for more canonical spin chain models is parametrically larger, of order $N$ or $N^2$.
This in turn implies that the SYK model has a much larger spectral rigidity than the most conventional spin chain ergodic models, \textit{i.e.} it is strongly chaotic.
We believe that the very large spectral rigidity of the SYK Hamiltonian is the key ingredient behind the excellent performance of the SYK QB.
As a concluding remark on this aspect, to further corroborate the agreement with RMT,  we can compare the SYK performance with the one of a QB in which the quench Hamiltonian is extracted directly from the Gaussian unitary ensemble (GUE), \textit{i.e.} in which the quench Hamiltonian is a random hermitian matrix, although such a QB does not represent any physical model by itself.
For the reasons just explained, \textit{i.e.} for the role played by the spectral rigidity, the GUE based QB would represents the upper limit to the possible charging stability of a generic QB, since the spectral rigidity of a GUE matrix is, by definition, maximal. This comparison is reported in Fig.~\ref{fig:GUE_vs_SYK}.
\begin{figure}[!t]
  \includegraphics[width=8cm]{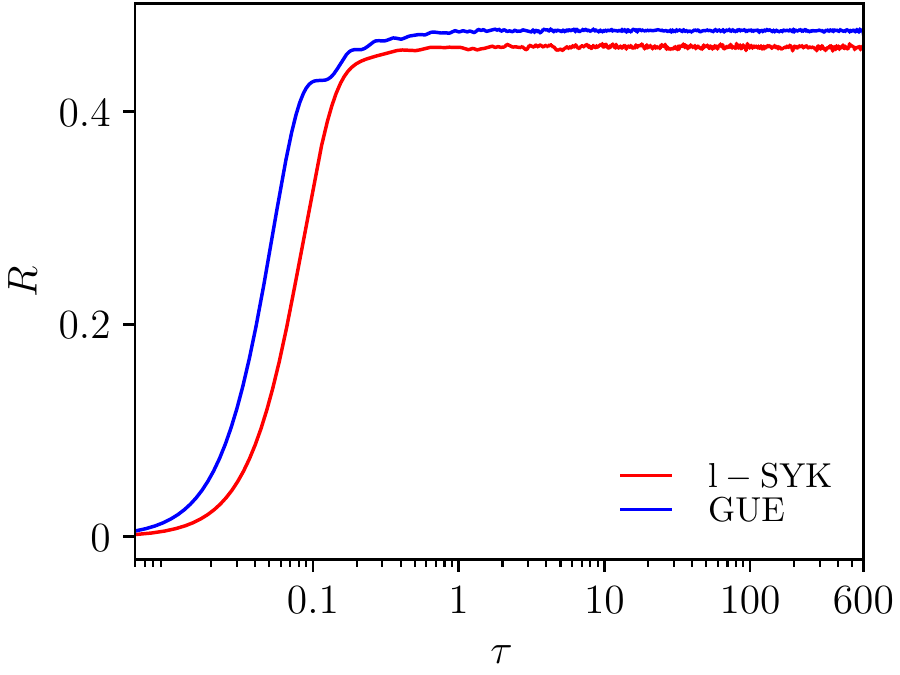}
   \caption{
    The charging ratio $R(\tau)$ —see Eq.~\eqref{eq:ratio_en_def}— as a function of $\tau$, for a single realization of the coupling constants $J_{ijkl}$, for both the quartic, local, SYK model, \eqref{eq:SYK_4_hamiltonian}, and a random Hamiltonian extracted from GUE.
   Both the models are computed for $N = 15$.}
  \label{fig:GUE_vs_SYK}
\end{figure}
Interestingly, we see that the performance of the SYK QB is very similar to the GUE QB, thus suggesting that the SYK QB, with its high level of spectral rigidity \cite{Gharibyan:2018jrp}, is likely to reach the upper bound on the possible charging stability of a generic, physical, QB.
}

\section{The ergotropy as a measure of thermalization}
\label{sec:ergotropy}

\begin{figure}[!t]
  \subfigure[\label{fig:figX_l-SYK}]{\includegraphics[width=8cm]{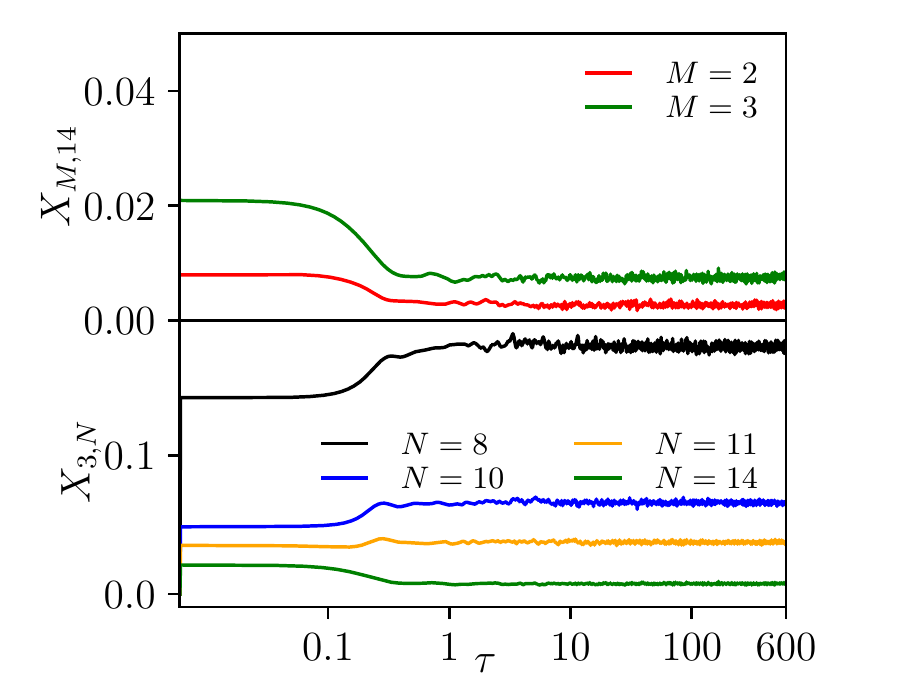}}
  \subfigure[\label{fig:figX_MBL}]{\includegraphics[width=8cm]{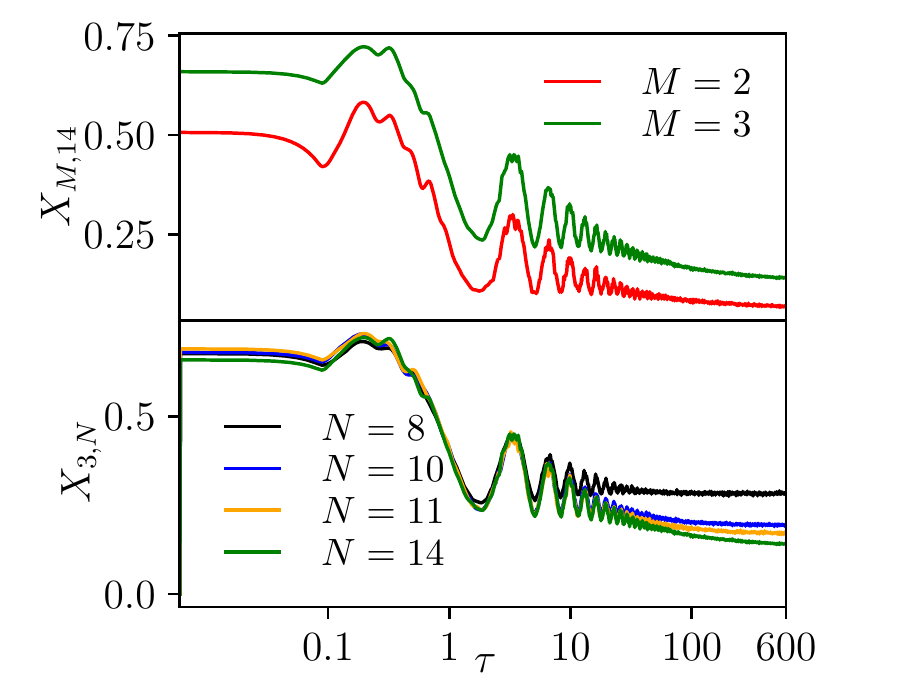}}
  \caption{The ratio $X_{M,N}(\tau)$, as defined in~\eqref{eq:ergotropy_def}, for a SYK battery~\subref{fig:figX_l-SYK}
    and for a MBL battery~\subref{fig:figX_MBL}.
    Upper parts of the two panels: we fix $N = 14$ cells and reduced to $M = 3$ or $M = 2$ cells;
    lower parts of the two panels: $N = 8, \, 10, \, 11, \, 14$ cells and reduced to $M = 3$ cells.
    Results are obtained by averaging over $750$ (up to $N =11$), $500$ ($N = 12 , \, 13$) and $200$ ($N = 14$) ensemble realizations.}
  \label{figX}
\end{figure}

Another important quantity, which characterizes the performance of a QB, is the so-called \emph{ergotropy}, $\mathcal E$~\cite{allahverdyan2013, andolina2019}.
Let us recall that it quantifies the amount of extractable work from a QB after the charging protocol~\cite{campaioli2018, allahverdyan2013, andolina2019}.
Indeed, if one assume to have access to just $M < N$ cells of the full QB, part of the energy stored will be locked by internal correlations, thus reducing the efficiency of the QB itself.
Given a density matrix $\rho$, representing the evolved state $\ket{\psi(\tau)}$ after tracing out the useless $N - M$ cells, the associated ergotropy is:
\be
\label{eq:ergotropy_def}
\mathcal{E}^{(N)}_M  \equiv \mathrm{Tr}[\mathcal{\hat H}_0^{(M)} \rho] - \mathrm{min}_{\hat U} \left\{ \mathrm{Tr}[\mathcal{\hat H}_0^{(M)} \, \hat U \rho \hat U^\dagger] \right\} \ ,
\ee
where $\mathcal{\hat H}_0^{(M)}$ denotes the local portion of the Hamiltonian, \eqref{eq:charging_protocol_general}, once restricted to the $M$ cells (we are assuming that $\mathcal{\hat H}_0$ can be written as a sum of local terms, such that it makes sense to define $\mathcal{\hat H}_0^{(M)}$) and the minimization runs over all the possible unitaries, $\hat U$, acting on $\rho$.

It is known that the ergotropy is severely affected by the presence of entanglement, \cite{andolina2019}: if the evolved state $\ket{\psi(\tau)}$ is highly entangled, the resulting density matrix $\rho$ will be highly mixed, and the corresponding levels of ergotropy will be very low, thus showing that, in this case, the amount of extractable energy from a subset of $M$ cells is low.
An interesting quantity to study is the following
\begin{equation}
  \label{eq:ErgoMN}
  X_{M,N}(\tau) \equiv \big\langle\!\big\langle \frac{\mathcal{E}^{(N)}_M (\tau)/M}{E_N (\tau)/N} \big\rangle\!\big\rangle \ ,
\end{equation}
which quantifies the fraction of energy, \emph{per cell}, that can be extracted from a reduced battery of $M$ cells, out of the initial $N$ cells.
It is particularly worth to study the behavior of $X_{M,N}(\tau)$, both as a function of $M$, at fixed $N$, and as a function of $N$, at $M$ fixed.

We have studied, for both the MBL and the SYK QBs, $X_{M,N}(\tau)$ for $N=14$ and very small values of $M$,
and also for $M=3$ and various values of $N$.
The results are depicted in Fig.~\ref{figX}.
From the upper panel of  Fig.~\ref{fig:figX_l-SYK} we see that $X_{M,14}(\tau)$, for the SYK battery, is very low.
This result follows from the fact that the SYK Hamiltonian is highly entangling, as discussed in Ref.~\cite{liu2017, huang2019}.
Moreover, from the lower panel of Fig.~\ref{fig:figX_l-SYK} we learn that the value of $X_{M,N}(\tau)$, at a given value of $M$, is highly affected by the size of the full battery, with $X_{3,N}(\tau)$ which decreases by increasing the dimension of the  battery.

Summarizing, the fact that Eq.~\eqref{eq:early_late_fluct_l-SYK} signals an exponential suppression of fluctuations, together with the observation that the ergotropy decreases with $N$, suggests that it is not convenient to build big batteries based on the SYK protocol and just keep a small portion of them at the end of the charging, while it is much more convenient to work directly with small batteries.

Moving to the MBL case, the situation is very different: from the upper panel of Fig.~\ref{fig:figX_MBL} we see that the amount of extractable energy is by far higher in agreement with the results of Ref.~\cite{rossini2019}, where it was observed that the levels of ergotropy for the MBL system are generally very high, a feature that can be traced back to the low level of entanglement typical of the MBL phase.
Much more interesting is the time behavior which can be observed in the lower panel of Fig.~\ref{fig:figX_MBL}: at early times we clearly see that the amount of extractable energy per cell is, essentially, \emph{independent} of the value of $N$, a behavior which is in striking contrast with what we observed for the SYK battery.
However, moving to later times, the situation changes and the amount of extractable energy per cell becomes \emph{$N$-dependent}, and in particular it gets reduced by increasing $N$, showing a behavior qualitatively similar to the SYK battery.

This observation confirms the picture we outlined in the previous section: the dynamics of the MBL battery shows a clear change when passing from early times to late times.
The behavior at early times is similar to the one expected for an integrable system, while at late times it becomes more similar to the behavior of a chaotic system.
The crossover between the two behaviors is in correspondence with the thermalization of the system and, once again, we stress that it should tend to infinity in the thermodynamic limit for the MBL system contrary to the SYK, for which the thermalization properties have been studied in the large $N$ limit, see Refs.~\cite{eberlein2017,bhattacharya2018}.

\section{Summary and outlook}
\label{sec:conclusions}
In this paper, we have introduced a new class of quantum batteries, in which the unitary charging protocol is realized via a sudden quench with a SYK-like Hamiltonian.
We have argued, and shown via extensive numerical computations, that such a charging protocol is able to dramatically suppress the strength of the temporal fluctuations.

As a byproduct, we have found evidence that a new interesting time scale can be uncovered during the charging of a quantum battery; namely the time scale at which the charging protocol turns to be collective, which corresponds to the time at which one can observe a transition in the strength of the temporal fluctuations as a function of the size of the system.
We have also provided a microscopic understanding of this new time scale, as the one at which an initially localized state (in the eigenbasis of the constant Hamiltonian) has spread to cover a large portion of the eigenbasis of the constant Hamiltonian.

By making use of this last point of view, and using also the temporal evolution of the ergotropy as a further probe, we then conjecture that the high stability of the charging protocol based on the SYK model is just another manifestation of the fast scrambling (and fast thermalizing) property of the SYK Hamiltonian, thus suggesting that the stability reached by the SYK quench puts an upper bound on the level of stability that a QB can show.

Of course, there are many open points which would be worth to explore.
It would be desirable to find further evidences for the conjecture that the charging stability of the SYK QBs is an upper bound for the charging stability of a generic QB.
{The results of our paper suggest an interesting connection between the charging stability and the degree of the spectral rigidity of a chaotic quench Hamiltonian, promoting the latter as another useful quantity to determine how strong is its chaotic behaviour and hence the related performance of a generic QB.}
Another promising line of research would be to study the charging protocol described in this paper from the holographic point of view, perhaps along the lines of \cite{dhar2018}.
Such an approach could be also relevant both to confirm the presence of an upper bound on the possible charging stability of a quantum battery and also to find its possible implications in the physics of the black holes.

\begin{acknowledgments}

D. Rosa and M. Carrega would like to thank M. Baggioli, F. Cavaliere, J. Kim, J. Murugan, T. Nosaka, and M. Poggi for stimulating discussions. D. Rosa also gratefully acknowledges the Simons Center for Geometry and Physics, Stony Brook University, and the organizers of the program ``Universality and ergodicity in quantum many-body systems'', for support and the warm hospitality. Some of the numerical computations  were performed  thanks to the Korea Institute for Advanced Study which provided computing resources (KIAS Center for Advanced Computation Abacus System). The work of D. Rosa is supported by a KIAS Individual
Grant PG059602 at Korea Institute for Advanced Study. M. Carrega acknowledges support from the Quant-Era project ``Supertop''.

\end{acknowledgments}

\appendix

\section{The role of the {battery} Hamiltonian $\mathcal{\hat H}_0$}
\label{app:l-SYKvsml-SYK}

An interesting check to perform is to investigate the role of the local term in the Hamiltonian, $\mathcal{\hat H}_0$, on the charging performance of a given QB.
To this end, we can study a slightly different version of the SYK model, usually called ``mass-deformed'' SYK model ($m$-SYK), studied in \cite{garcia-garcia2017,nosaka2018,Gharibyan:2018jrp}.
In this model, the quench Hamiltonian, $\mathcal{\hat H}_1$, is the usual quartic Hamiltonian of the SYK model, as defined in \eqref{eq:SYK_4_hamiltonian}, while the constant term is given by the nonlocal random mass term, defined in \eqref{eq:SYK_2_hamiltonian}.

\begin{figure}[!t]
  \includegraphics[width=8cm]{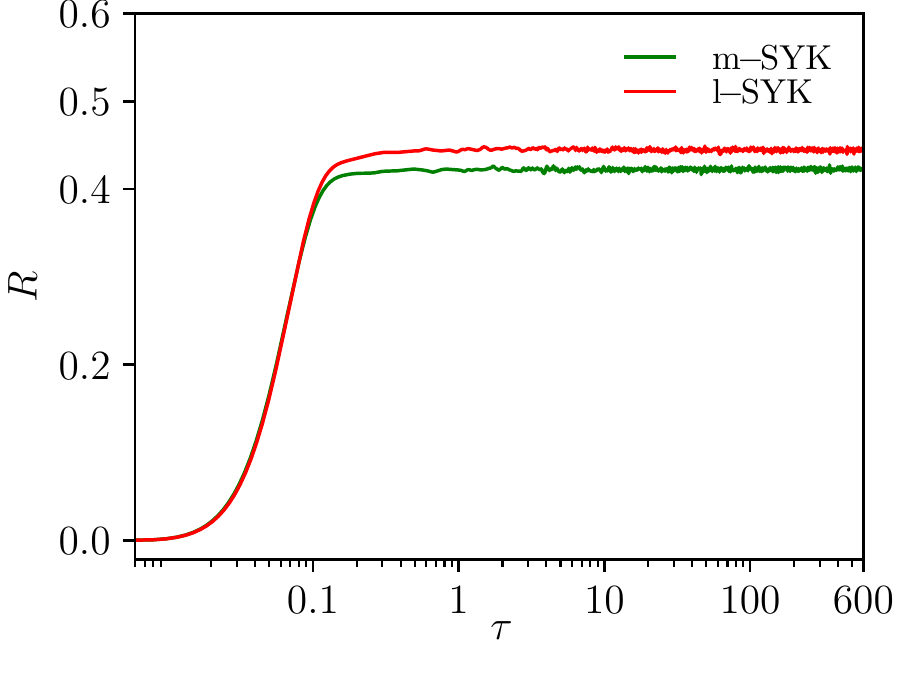}
  \caption{The charging ratio as a function of time $\tau$ (in units of $h$) for both the SYK and the $m$-SYK batteries, for a battery of length $N = 15$ and for a single realization of the disorder random couplings.}
  \label{fig_app_mass}
\end{figure}

We have compared, for the same realization of the disorder couplings $J_{ijkl}$ in both the models and for a realization of $K_{ij}$, the function $R(\tau)$ for both the SYK and the $m$-SYK batteries.
We have renormalized the bandwidth of $\mathcal{\hat H}_2$ such that the constraint $\Delta_{\mathcal{\hat H}_0} = \Delta_{\mathcal{\hat H}_2}$ was satisfied.

From Fig.~\ref{fig_app_mass} we clearly see that the two performances are almost the same, both in terms of the maximal value reached by $R(\tau) $, and in terms of the strength of the fluctuations, with a small advantage for the SYK model.
This shows that the role of the particular $\mathcal{\hat H}_0$ term on the charging performance is very limited, and that only the quench Hamiltonian really matters in the unitary charging protocol.

\section{Temporal fluctuations in the AL phase}
\label{app:And_time}

\begin{figure}[!t]
  \includegraphics[width=8cm]{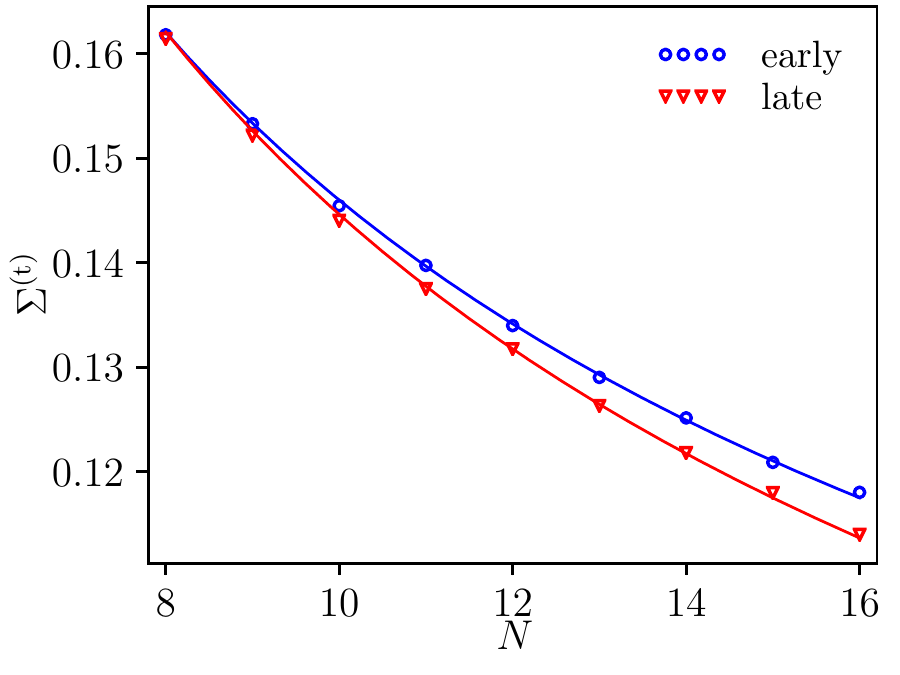}
  \caption{The early and late time fluctuations, as measured by $\Sigma^{(\rm t)}$, for the Anderson spin chain, as a function of the lattice size. Continuous lines correspond to the fit in Eq.~\eqref{eq:early_fluct_MBL}. The results have been obtained by averaging over $500$ ensemble realizations.}
  \label{fig5and}
\end{figure}

For completeness, we report in Fig.~\ref{fig5and} the behaviour of the temporal fluctuations for the Anderson model,
analogous to the plots discussed in Figs.~\ref{fig:MBL_fluctuations}-\ref{fig:l-SYK_fluctuations}.
We see that, in this case, in both time windows (early and late time behaviours), the data are greatly reproduced
by a $\frac{1}{\sqrt N}$-like function.
This result is in line with our expectation, since the Anderson model does not thermalize and, as discussed in Fig.~\ref{fig1}, it shows huge temporal fluctuations at all the time scales, contrary to what happens for the other two models considered.
This implies that in the AL phase a collective behaviour, with a corresponding exponential suppression of temporal fluctuations, is never reached even at late times.

{
\section{Chaotic properties and QB performance}
\subsection{Chaotic properties of the quadratic SYK Hamiltonians}
\label{subsec:quantum_chaos_SYK}

In this section, we show that, despite being very similar, the two quadratic, SYK-like, Hamiltonians \eqref{eq:SYK_2_hamiltonian} and \eqref{eq:SYK_2_hamiltonian_bosonic} present very different properties, with the fermionic Hamiltonian being integrable and the bosonic Hamiltonian being chaotic.

To this end, {\it i.e.} to show  the chaotic/integrable nature of the two models, it is sufficient  to focus on a short-range diagnostics of quantum chaos, \textit{i.e.} testing the agreement with the RMT preditions for very small energy separations, of order of the mean level spacing.
In particular, we consider the so-called r-statistics, also known as adjacent gap ratio \cite{Oganesyan_2007}.
This quantity, which is equivalent to the well-known nearest neighbor spacing distribution defined, for example, in \cite{Guhr:1997ve}, has the advantage that it does not require to rescale the spacings by the mean level density.
In other words, it does not require an unfolding of the spectrum, that can be a delicate issue \cite{Oganesyan_2007}.

The r-statistics can be operatively defined as follows:
For a given Hamiltonian spectrum, the distinct energy levels, $E_i$, are listed in ascending order
\begin{equation}
  E_1 \leq E_2 \leq E3 \dots \ ,
\end{equation}
and the corresponding nearest level spacings are computed as
\begin{equation}
  \delta_i \equiv E_{i+1} - E_i \ .
\end{equation}
Finally, one has the ratios
\begin{equation}
  r_i \equiv \frac{\mathrm{min}(r_i , r_{i + 1})}{\mathrm{max}(r_i , r_{i + 1})} \ .
\end{equation}

The ratios $r_i$, once averaged over many ensemble realizations, can be used as a short-range diagnostics of quantum chaos.
Indeed, it can be shown \cite{r-stat_second} that for a non-chaotic model, the average values $\langle\langle r_i \rangle\rangle$ agree with the predictions for a Poissonian  spectrum, $\langle\langle r_i \rangle\rangle \sim 0.386$. On the other hand, for a chaotic spectrum, the values of $\langle\langle r_i \rangle\rangle$ are larger than $0.5$ and, more precisely, they agree with the predictions of RMT, which are $\langle\langle r_i \rangle\rangle  \sim 0.536$, $\langle\langle r_i \rangle\rangle  \sim 0.603$ and $\langle\langle r_i \rangle\rangle  \sim 0.676$ for the Gaussian orthogonal ensemble (GOE), Gaussian unitary ensemble (GUE) and Gaussian symplectic ensemble (GSE), respectively.

Given these preliminaries, we have computed for both the Hamiltonians \eqref{eq:SYK_2_hamiltonian} and \eqref{eq:SYK_2_hamiltonian_bosonic} the r-statistics. 
\begin{figure}[!t]
  \includegraphics[width=8cm]{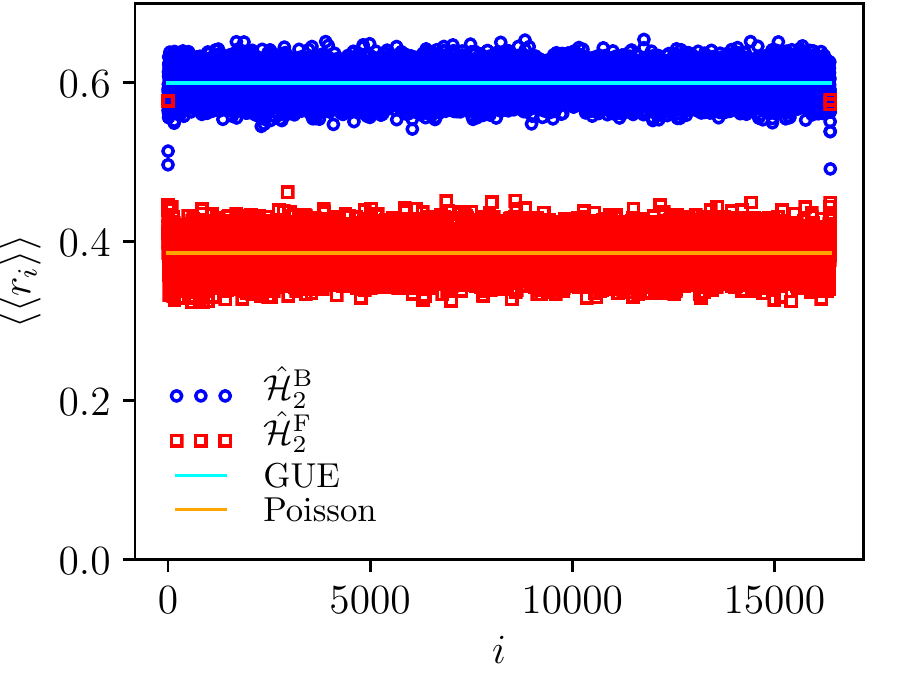}
  \caption{The values of the r-statistics, $\langle \langle r_i \rangle \rangle$, for both the fermionic quadratic model, $\mathcal{\hat H}_2^\mathrm{F}$, and the bosonic model as well, $\mathcal{\hat H}_2^\mathrm{B}$, for $N = 15$.
  The prediction of GUE and Poisson are also reported.
  The values are averaged over $700$ ensemble realizations.}
  \label{fig:r-statistics}
\end{figure}
The results are reported in Fig.~\ref{fig:r-statistics}, where we clearly see that the quadratic fermionic Hamiltonian shows integrable behavior, while the bosonic Hamiltonian is clearly chaotic.

\subsection{The charging stability for an ergodic spin chain}
\label{subsec:ergodic_vs_SYK}

One may wonder whether a more conventional chaotic Hamiltonian, instead of the SYK model, like the one in \eqref{eq:QBham1_MBL} in the ergodic phase (obtained by setting $J = \delta J = 1.67h$ and $J_2 = 0.5 h$), can be as efficient as the SYK Hamiltonian in reducing the temporal fluctuations.

In Fig.~\ref{fig:ergodic_vs_SYK} we compare the charging protocol between the ergodic spin chain \eqref{eq:QBham1_MBL} and the SYK protocol, for $N = 15$ cells.
\begin{figure}[!t]
  \includegraphics[width=8cm]{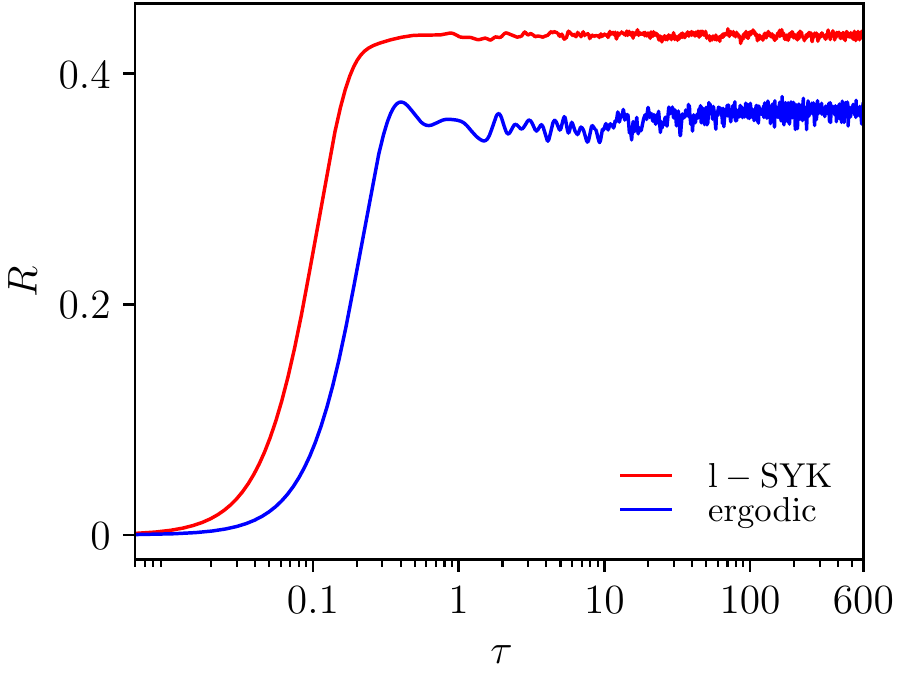}
   \caption{The charging ratio $R(\tau)$ —see Eq.~\eqref{eq:ratio_en_def}— as a function of $\tau$, for a single realization of the coupling constants, for both the quartic local SYK model, \eqref{eq:SYK_4_hamiltonian}, and the spin chain model, \eqref{eq:QBham1_MBL}, in the ergodic phase.}
  \label{fig:ergodic_vs_SYK}
\end{figure}
It is immediate to see that the SYK QB is much more efficient in suppressing the temporal fluctuations, thus showing that quantum chaos, solely determined looking at the RMT predictions of the short-range diagnostics of chaos, does not guarantee the same quality of the SYK QB.
We can study also for the ergodic spin chain the suppression of the temporal fluctuation with the number of cells $N$, \textit{i.e.} the behavior of $\Sigma_N^{(t)}$, as done for the MBL and the SYK models in \eqref{eq:early_fluct_MBL} -- \eqref{eq:early_late_fluct_l-SYK}, respectively. 
\begin{figure}[!t]
  \includegraphics[width=8cm]{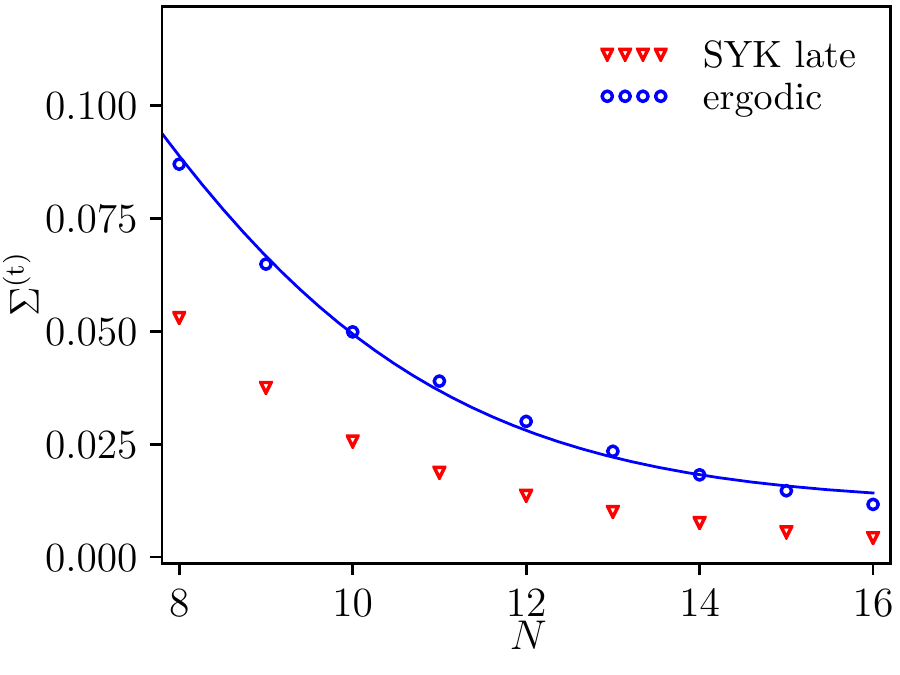}
  \caption{The late time fluctuations, as measured by $\Sigma^{(\rm t)}$, for the ergodic spin chain, as a function of the lattice size. The continuous line corresponds to the fit $\Sigma^{(\rm t)} \equiv a N^3 2^{- N} + b$, with $a$ and $b$ being fitting parameters. The results have been obtained by averaging over $500$ ensemble realizations.
  For comparison, the values of $\Sigma^{(\rm t)}$ for the l-SYK battery are also reported.}
  \label{fig:late_fluct_ergodic}
\end{figure}
As already noticed for the SYK QB, also in this case the fluctuations at early and at late time are equivalent, and more precisely, as we report in Fig.~\ref{fig:late_fluct_ergodic}, we see that the fluctuations are again exponentially truncated by increasing the system size, but by comparing their strength with the SYK fluctuations, as in Fig.~\ref{fig:l-SYK_fluctuations} (and reported here for convenience), we see that they are significantly larger than in the SYK case.
This suggests that the exponential suppression of the temporal fluctuations is a generic feature of quantum chaos, as diagnosed by the short-range diagnostics, but that the strength of the suppression with $N$ is controlled by the level of spectral rigidity.}

\end{document}